 \documentclass[12pt]{iopart}
 \usepackage{iopams}
 \usepackage{epsfig}   
 \usepackage{graphics}
 \usepackage[T1]{fontenc}

 \renewcommand{\sf}{\sffamily}
 \renewcommand{\bf}{\bfseries}
 \renewcommand{\it}{\itshape}

 \newcommand{\nn}{\nonumber}
 \newcommand{\er}[1]{(\ref{#1})}
 
 \renewcommand{\d}{{\rm d}}     
 
 \newcommand{\td}[2]{\frac{{\rm d} {#1}}{{\rm d} {#2}}}
 \newcommand{\tdil}[2]{{\rm d} {#1} / {\rm d} {#2}}
 \newcommand{\tdt}[2]{\frac{{\rm d}^2 {#1}}{{\rm d} {#2}^2}}
 \newcommand{\tdtil}[2]{{\rm d}^2 {#1} / {\rm d} {#2}^2}

 \newcommand{\LRa}{\Leftrightarrow}

 \newcommand{\LT}{Lema\^{\i}tre-Tolman}
 \renewcommand{\L}{Lema\^{\i}tre}
 
 \newcommand{\LCDM}{$\Lambda$CDM}

 \newcommand{\Rt}{\dot{R}}
 \newcommand{\Rtt}{\ddot{R}}
 \newcommand{\Rh}{\widehat{R}}
 \newcommand{\Rth}{\widehat{\dot{R}\,}}
 \newcommand{\Rrh}{\widehat{R'\,}}
 \newcommand{\Rtth}{\widehat{\ddot{R}\,}}
 \newcommand{\Rtrh}{\widehat{\dot{R}'\,}}
 
 \renewcommand{\th}{\widehat{\,t}}
 \newcommand{\rh}{\hat{r}}
 \newcommand{\Ut}{\dot{U}}
 
 \newcommand{\Uh}{\hat{U}}
 \newcommand{\Uth}{\hat{\dot{U}}}

\begin{document}
 
 \title{The Metric of the Cosmos from Luminosity and Age Data}

\author{Krzysztof Bolejko$^{1,2}$,
Charles Hellaby$^{2}$,
Alnadhief H. A. Alfedeel$^{2}$}

\address{$^{1}$ Astrophysics Department,
University of Oxford, Oxford OX1 3RH, UK}
\address{$^{2}$ 
Astrophysics, Cosmology and Gravity Centre (ACGC), 
Department of Mathematics and Applied Mathematics,
University of Cape Town, Rondebosch 7701, South Africa}
 \ead{\mailto{Krzysztof.Bolejko@astro.ox.ac.uk}, \mailto{Charles.Hellaby@uct.ac.za},
 \mailto{Alnadhief.Alfedeel@uct.ac.za}}

 \begin{abstract}
 This paper presents the algorithm for determining the \LT\ model that best fits given datasets for maximum stellar ages, and SNIa luminosities, both as functions of redshift.  It then applies it to current cosmological data.  Special attention must be given to the handling of the origin, and the region of the maximum diameter distances.  As with a previous combination of datasets (galaxy number counts and luminosity distances versus redshift), there are relationships that must hold at the region of the maximum diameter distance, which are unlikely to be obeyed exactly by real data.  We show how to make corrections that enable a self-consistent solution to be found.  We address the questions of the best way to approximate discrete data with smooth functions, and how to estimate the uncertainties of the output --- the 3 free functions that determine a specific \LT metric.  While current data does not permit any confidence in our results, we show that the method works well, and reasonable \LT\ models do fit with or without a cosmological constant.
 \end{abstract}

 \section{Introduction}
  
   Over the past century, the investigation of the structure and evolution of the universe has become a major scientific thrust.  Einstein's equations made the geometry of the cosmos inseparable from its matter content.  Thus there are many intriguing investigations,  geometric as well as physical, ranging from whether the spatial section of universe is open or closed, and what the global topology is, down to lensing by individual mass concentrations, and detecting the signatures of galactic black holes.  Equally important is the geometry on a range of intermediate scales, and its effect on the paths of the light rays that constitute the bulk of our cosmological observations.
   
The assumption of a homogeneous Friedmann-\L-Robertson-Walker (FLRW) universe made the Einstein equations tractable, and later, complemented by perturbation theory, opened the way to our present day understanding of an observable universe that is $10^{26}$ times larger than a human frame, and $10^8$ times older than a human lifespan.  
Currently, in most cases the cosmological observations are anlysyed within the 
 framework of homogenous models, for example supernova observations \cite{snd} galaxy redshift surveys \cite{sdss} in particular baryon acoutiong oscilations \cite{bao}, and cosmic microvawve backrogund radition \cite{wmap7}.  The cosmological principle remains a core ingredient of the standard concordance model, but as our measurements become ever more complete and detailed, deviations from homogeneity will have to be taken seriously.  Certainly, inhomogeneity exists on a wide range of scales, and a knowledge of the associated spacetime curvature is indispensable to the detailed analysis of observational data.  

Whether one views it as detecting what level of inhomogeneity exists, or verifying how well homogeneity is satisfied, the investigation is essential.  The most important reason is that the scientific method requires it --- if you {\it can} test an assumption you {\it should} --- not just in one way, but in many different ways.

The history of science can provide many examples of why different tests are important.  The most well known example is probably the Ptolemaic vs Copernican model of the Universe.  At the time of Nicolaus Copernicus, the Ptolemaic model with its large number of epicycles was able to fit cosmological/astronomical data much better than the Copernican model.  Thus, we should bear in mind that consistency with 
current observations is not a final 
{\it proof} of the underlying assumptions of the model, and as more data become available we should test our assumptions using different methods.  

One  way of testing the assumption of homogeneity is to look
 at the consistency relations between the observables.
 A  consistency check is based on  verifying whether
 the relation between observations is as given by the homogeneous cosmological model. 
 Previously,  Ribeiro and Stoeger considered the consistency between
 the expected number counts in a homogeneous model, a commonly used 
 galaxy luminosity function, and  the observed
 galaxy number counts  \cite{RiSt2003}. Later they showed that such an analysis strongly depends on the
distance definition used \cite{AIRS2007}. In \cite{ClBH2008} Clarkson, Bassett \& Lu
showed that in Friedmann models that
$H(z)$ and $D(z)$ are not independent but must obey a special relation.
Thus, by checking if observations obey  this
 relation, one can test the assumption
of homogeneity. A limited version of such a test was proposed in \cite{SaSS2008},
where just $H(z)$ alone can be used to test the consistency between observations
and a spatially flat homogeneous model.
Another interesting aspect is to check the consistency between the age of the universe
inferred from the cosmological models with `local' (age of meteorites or
stars in our own Galaxy) measurements.

In this paper we take another approach.
We study the observations within an inhomogeneous framework.
The motivation is as follows: if the the Universe is indeed homogeneous, then
the data should favour a homogeneous model over inhomogeneous ones.

As argued in \cite{BKHC2009}, spherically symmetric models are the logical first step away from homogeneity.  Radial inhomogeneity is intrinsically hard to detect and separate from other past null cone effects.  Our observations reveal a tilted slice through spacetime, and what we see is affected by the details of the cosmic evolution due to the bulk equations of state, the evolution of light sources, and the lumpiness of the matter distribution.

The simplest spherically symmetric, non-stationary, 
non-homogeneous solution of the Einstein equations is the \LT\ (LT) model.  Currently there is a large number of papers that use the LT model to fit cosmological data without dark energy --- a non-compete list includes
 \cite{Cel00, IgNN2002, AAG06, chung06, AlAm2007, EnqMat07, BMN07, BTT07a, BTT07b, MKMR2007, bolejko08, MK2008, BT08, GH08a, GH08b, CFL08, ClFZ2009, Enqv08, BN08, ZaMS2008, BoWy2009, EMR09, ABNV2009, WV09, RC10, ClRe2010, MZS10, LLLW10, ZhSt2010, YNS10P1, YNS10P2, NaSa2010,DGOU2010}; for a review see \cite{BoCK2011,MaNo2011}.  
However, these methods are based on what is often called the {\it direct approach}, which can be summarized as follows: `assume a model, fit the model to the data, obtain the best fit parameters of the model and compare the 
 goodness of fit  with the standard cosmological model'.
The caveat of this kind of approach is that any assumed parametrization is not general, and may omit a large part of the set of all possible models, hence a better approach is the so called {\it inverse approach}.
Within the inverse approach one uses the data to construct the model.  
In theory, this approach 
is much less restricted by choice of model and parameterization than 
the direct approach.  In practice however one still needs to impose {\em some} additional assumptions to make the calculations tractable.  
The problem is that we need to be able to solve the Einstein equations.  Thus, one either employs a series expansion around the observer \cite{KrSa1966,PaMa1984,PaSa1999}, or assumes spherical symmetry to calculate the evolution of the Universe within the whole observable domain.  Here one can either use 
`observer coordinates' and the fluid-ray tetrad approach \cite{EN85, StEN1992, MHMS96, AABFS01, HeAl2009}, or use standard coordinates in the spherically symmetric dust model, i.e. the LT model, and determine the null-cone data relations
\cite{MuHeEl97, Hel06, LH07, MH08, YoKN2008, CeBK2010, Chul2010, Roma2010JCAPa, Roma2010JCAPb, Roma2010PRD}.

 In this paper we will use the inverse method with the \LT\ model to test the assumption of homogeneity.  
The motivation and plan is as follows: 
`start with an inhomogeneous model more general than FLRW, apply the inverse method to 
the actual observational data to specify the model, then if the model obtained in this way is homogeneous (or very close to homogeneous) we 
 find in favour of homogeneity, if on the other hand the model is significantly
inhomogeneous, we disfavor the homogeneity'.  It should be emphasised that we can only {\it favour} or {\it disfavor}, not actually {\it prove} or {\it rule out}.  As shown in Ref. \cite{MuHeEl97} a \LT\ model could be found to fit any reasonable observational data for the luminosity (or diameter) distance versus redshift, and the galaxy number-counts versus redshift; while the difficulty of separating cosmic evolution, source evolution, and inhomogeneity was pointed out.  This concept was expanded, with an important correction, and explicit numerical implementations were developed in \cite{Hel06,LH07,MH08}.  
A possible method for testing source evolution theories in the presence of inhomogeneity was suggested in \cite{Hel01}.

It is important to begin making the connection with real data, as we do here.  We note however, that the quantity and quality of current data is not sufficient to draw any clear conclusions from the LT models obtained.  The purpose is to explore the various options and uncover potential difficulties, so that a proper analysis of future data will be possible.

It is also important to investigate the question of homogeneity using many different methods and datasets.  Consequently, this paper applies the metric of the cosmos approach to a different set of observations.  It uses the luminosity distances of supernovae and the maximum ages of galaxies, both as functions of redshift.  This is in contrast to using the redshift variations of luminosity or diameter distances and source number counts, as considered previously \cite{Hel03,LH07,MH08}.

 \section{The \LT\ solution}
 \label{lt}

   The LT (\LT) model \cite{Lem33,Tol34} is a spherically symmetric solution of Einstein's equations where the
gravitational source is dust. In comoving and synchronous coordinates, the metric is
 \begin{equation}\label{LTmet}
  \d s^2 = - \d t^2 + \frac {R'^2}{W^2(r)} \d r^2 +
   R^2(t,r)(\d \theta^2 + \sin^2\theta \d \phi^2) ~,
 \end{equation}
 where $W(r) = 1 + f(r)$  
is an arbitrary function, $R' = \partial R/ \partial r$.  The spatial sections are flat only if $W(r) = 1$, so $W$ is a geometric factor.  The areal radius $R(t,r)$ obeys
 \begin{equation}\label{vel}
   \Rt^2 = W^2 -1 + \frac{2M}{R} + \frac{\Lambda}{3} R^2 ~,
 \end{equation}
 where $\Rt = \partial R / \partial t$ and $\Lambda$ is the cosmological constant.  From \er{vel}, $E = (W^2 - 1)/2$ is the total energy per unit mass of the dust particles.  Equation \er{vel} is a first integral of the Einstein equations, and $M(r)$ is another
arbitrary function of integration that gives the gravitational mass within each comoving shell of coordinate radius $r$.  The mass density in energy units is:
 \begin{equation}   \label{den}
   \kappa \rho = \frac {2 M'}{R^2 R'} ~, 
 \end{equation}
 where $\kappa = {8 \pi G}/c^4$.  Equation \er{vel} can be solved by simple integration
giving a relation for the age of the Universe, $\tau$
 \begin{equation}\label{tbf}
   \tau(r) = t- t_B(r) = \int\limits_0^R\frac{{\rm d} \widetilde{R}}{\sqrt{W^2 -1 + 2M/\widetilde{R} +
   \frac{1}{3}\Lambda \widetilde{R}^2}} ~,
 \end{equation}
 where $t_B(r)$ appears as an integration function, and is an arbitrary function of $r$ giving the time at which $R = 0$ on each worldline.  This means that the Big Bang is not simultaneous as in the FLRW models, but occurs at different times at different distances from the origin.   Since all the formulae given so far are covariant under coordinate transformations of the form $\widetilde{r} = g(r)$, one of the functions $W(r)$, $M(r)$ and $t_B(r)$ can be fixed at will by the choice of $g$.%
 \footnote{But note that no one choice can cover all possible LT models; any given $M(r)$ either does or does not have a vacuum region, it either does or does not have a spatial maximum; any $f(r)$ either does or does not change sign; any $t_B(r)$ either does or does not have a constant region.  
 }%
Therefore, once this choice is made, a given L--T model is fully determined by two of these arbitrary functions.  In other words, we need two observed relationships to pin down an LT model.  For more information on the LT model, see \cite{Kra97,PlKr06,Hel09}.

 \subsection{Past Null Cone and Observations}
 \label{nco}

   Since light propagates on null geodesics, let us define {\em the} past null cone (PNC) by the observation event $(t, r) = (t_0, 0)$ for a central observer at the present day.%
 \footnote{As stated elsewhere, there are good reasons for working initially with a central observer: much present-day data is averaged over the whole sky just to get a big enough dataset, we do live at the centre of our own past null cone, and radial inhomogeneity is much harder to detect than anisotropy. 
 There is no  implicit claim we live at the centre of a gigantic inhomogeneity.}
 The equation for incoming radial null geodesics is
 \begin{equation}
   \td{t}{r} = \frac{- R'}{W} ~.
 \end{equation}
 Let us choose the radial coordinate so that, on the PNC,
 \begin{equation}
   \frac{\Rrh}{W} = 1 ~, \label{corchoice}
 \end{equation}
 where by a hat or a subscript wedge (e.g. $\hat{Q}$ or $Q_\wedge$) we denote any quantity evaluated on the PNC.  (Note: this choice of $r$ is possible only on a single light cone.  In the following, we always refer to the PNC of  $(t_0, 0)$.)  This  gauge choice 
simplifies the  PNC equation:
 \begin{equation}
   \th(r) = t_0 - r ~.
   \label{ngs}
 \end{equation}
 The equation for the redshift reads \cite{Bon47,MuHeEl97}
 \begin{equation}
   \frac{1}{1 + z(r)} \frac{{\rm d}z(r)}{{\rm d}r} = \frac{\Rtrh}{W} ~. \label{dz}
 \end{equation}

 We aim to show how to define an LT model from luminosity and age data,
 so we need to express the null cone relations using observables like luminosity (or diameter) distance 
and the age of the Universe.
  In the LT model with observer at the origin, the angular diameter distance equals the areal radius $\Rh$ at the point of emission.  Using the reciprocity theorem \cite{Eth33,Pen66,E1971} this can be related to the luminosity distance:  
 \begin{equation}
   \Rh = D_A = \frac{D_L}{(1+z)^2} ~.
 \end{equation}

   Taking the total derivative of the areal radius $\Rh$ gives \cite{MuHeEl97,Hel09}
 \begin{equation}
   \td{\Rh}{r} = \Rrh + \Rth \, \td{\th}{r} = W - \Rth ~,
   \label{dRdr}
 \end{equation}
 and
 \begin{equation}
   \fl
   \tdt{\Rh}{r} = \left[ \, \left( 1 - \frac{\Rt}{W} \right) \left\{ R'' - \Rt'
      + \left( \frac{\Lambda R}{3} - \frac{M}{R^2} \right) \right\}
      - \Rt' \left( 1 - \frac{\Rt^2}{W^2} \right) - \frac{M' \Rt}{W^2 R} \, \right]_\wedge ~.
      \label{Rhrr}
 \end{equation}
 Note that there is a sign error in eq.\ (3.19) of \cite{Hel09}.  However, what we obtain from observations are $\Rh(z)$ and $\tau(z)$, so we thus convert the above to derivatives with respect to redshift$z$, using
 \begin{eqnarray}
   \td{\Rh}{r} = \td{\Rh}{z} \frac{1}{\varphi} =  (W - \Rth) \frac{1}{\varphi}   \label{Rhz} ~, \\
   \td{^2 \Rh}{r^2} = \td{^2\Rh}{z^2} \frac{1}{\varphi^2} -  \td{\Rh}{z} \frac{1}{\varphi^3} \td{\varphi}{z} ~,
 \end{eqnarray}
 where 
 \begin{equation}\label{phi}
   \varphi = \td{r}{z} ~,
 \end{equation}
 and the redshift derivative of $\varphi$ is then
 \begin{equation}
   \td{\varphi}{z}  = \varphi \left( \frac{1}{(1 + z)}
      + \frac{\frac{\varphi}{W \Rh} \td{M}{z} + \tdt{\Rh}{z}}{\td{\Rh}{z}} \right) ~,   \label{phiz1} 
 \end{equation}
 which  can be expressed as
 \begin{eqnarray}
   && \td{\varphi}{z}  = \varphi \Bigg\{ \frac{\tdt{\Rh}{z}}{\td{\Rh}{z}} 
      + \frac{1}{\left( A - \frac{B}{\Rh} \right)} \Bigg[ - \frac{\varphi \td{\tau}{z}}{W \Rh \td{\Rh}{z}} +
      \frac{\varphi^2}{W \Rh \left( W \varphi - \td{\Rh}{z} \right)} \nn \\
   &&~~~~~~~~~~~~ + \frac{A}{(1 + z)}
      - \frac{\varphi B}{\Rh \td{\Rh}{z}} \left\{ \left( \frac{M}{\Rh^2} - \frac{\Lambda \Rh}{3} \right) \varphi
      + \frac{W}{(1 + z)} \right\} \Bigg] \Bigg\} ~,
      \label{phiz}
 \end{eqnarray}
 where
 \begin{equation}
   A = \int_0^{\Rh} \frac{\d R}{R \Rt^3}, \quad B = \int_0^{\Rh} \frac{\d R}{\Rt^3} ~.
 \end{equation}

   The second set of observations, the age measurements are given in the form of $\tau(z)$:
 \begin{equation}
   \tau(z)  = t(z) - t_B(z) = t_0 - r(z) - t_B(r(z)) ~.
   \label{tauf}
 \end{equation}
 The derivatives of $\tau$ with respect to redshift are
 \begin{eqnarray}
   \td{\tau}{z}  = \frac{1}{\Rth} \td{\Rh}{z} - \td{M}{z} A - W \td{W}{z} B ~.
   \label{tauz} 
 \end{eqnarray}

   Solving \er{dRdr} and \er{vel} for $W$ we get 
 \begin{equation}
   W  = \frac{1}{2} \left( \frac{d\Rh}{dr} \right)
      + \frac{\left( 1 - \frac{2 M}{\Rh} - \frac{\Lambda \Rh^2}{3} \right)}{2 \left( \frac{d\Rh}{dr} \right)} ~, 
   \label{Weq}
 \end{equation}
 which in terms of redshift is 
 \begin{equation}
   W = \frac{1}{2 \varphi} \left( \td{\Rh}{z} \right)
      + \frac{\left( 1 - \frac{2 M}{\Rh} - \frac{\Lambda \Rh^2}{3} \right) \varphi}{2 \left( \td{\Rh}{z} \right)} ~.
      \label{Wz} 
 \end{equation}
 Differentiating the above with respect to redshift, and using \er{tauz} to eliminate $\tdil{W}{z}$, we get:
 \begin{equation}\label{dmz}
   \td{M}{z} = \frac{ - \td{\tau}{z} + \frac{1}{\Rth} \td{R}{z} + \varphi W B \Rtth - \frac{W B \Rth}{1+z}}{A - \frac{B}{\Rh}} ~.
   \label{Mz}
 \end{equation}

   The arbitrary LT functions that are determined by the given $\tau(z)$ and $D_L(z)$ (or equivalently by $\Rh(z)$), can be calculated in the following way.  We first solve \er{Mz} for $M(z)$ [where $\Rth$ is given by \er{vel}, $\Rtth = -M/\Rh^2 + \Rh \Lambda/3$, and $\varphi$ from \er{Rhz} is $({\rm d} \Rh/ {\rm d} z)/(W - \Rth)$] to find $M(z)$ for the next step, i.e. $M_{i+1}$. Then $W_{i+1}$ follows from \er{tbf}.  Next $r(z)$ is evaluated by solving \er{phi}, and then $t_B(z)$ follows from \er{tauf}.  It is convenient to let $t_B(0) = 0$ so that $t_0 = \tau(0)$.  While the above are the basis for a numerical algorithm over most of the $z$ range, two loci  require special attention, the origin, and the maximum in $\Rh$(z) \cite{LH07}.

 \subsection{Origin}
 \label{orig}

   At the origin, $R(t, 0) = 0$ $\forall~t$, and similarly $\Rt(t, 0) = 0 = \Rtt(t, 0)$ etc.  
Regularity \cite{HelLak85,MusHel01} requires $M \sim R^3$ and $f \sim R^2$ as $r \to 0$ (where 
$f = W^2 - 1 = 2 E$).  It is then evident  that $W \to 1$ and from \er{corchoice} and \er{dRdr} 
that $\Rrh \to 1$  and $\tdil{\Rh}{r} \to 1$ here, meaning $R \sim r$.  Equation \er{vel} shows 
that $\Rt \sim r$ and $\Rt' \sim r^0$, so it follows from \er{dz} that $\tdil{z}{r} \sim \Rtrh$.  
Thus, with $z = 0$ here, several quantities in the above equations go to zero and we have 0/0, 
which makes a direct numerical solution difficult at this locus.  In order to find the solution 
in the vicinity of the origin, we use a Taylor series approach, and we write
 \begin{equation} 
\fl
   \rh = \sum_{i=1}^\infty \rh_i z^i ~,\quad
   \Rh = \sum_{i=1}^\infty \Rh_i z^i ~,\quad
   \tau = \sum_{i=0}^\infty \tau_i z^i ~,\quad
   M = \sum_{i=3}^\infty M_i z^i ~,\quad
   f = \sum_{i=2}^\infty f_i z^i ~,
   \label{OrigSeries}
 \end{equation}
 and of course $\varphi = \tdil{\rh}{z}$ etc.  Since there is a well-defined mapping between $r$ \& $z$ near the observer, we write
 \begin{equation}
   R = z U ~,~~~~\mbox{and}~~~~ \Rt = z \Ut ~.   \label{RzU}
 \end{equation}
 so that
 \begin{equation}
   \tau  = \int_0^{\Uh} \frac{dU}{\sqrt{\frac{2 M}{z^3 U} + \frac{f}{z^2} + \frac{\Lambda U^2}{3}}\;} ~.
 \end{equation}
 Inserting \er{OrigSeries} and re-expanding the integrand as a Taylor series gives%
 \footnote{Of course, \er{tauorigseries} does not constitute a Taylor series in $z$ for $\tau$, since the upper limit of each integral itself depends on $z$.  Rather, \er{tau0} and \er{tau1} constitute the first two coefficients of the Taylor series for $\tau$.}
 \begin{eqnarray}
   \tau  & \approx & \int_0^{\Uh} \Bigg\{ 1 - \left( \frac{M_4}{U} + \frac{f_3}{2} \right) \frac{z}{\Ut_0^2} \nn \\
   &&~~~~ - \left[ \left( \frac{M_5}{U} + \frac{f_4}{2} \right) \frac{1}{\Ut_0^2}
      - \left( \frac{M_4}{U} + \frac{f_3}{2} \right)^2 \frac{3}{2 \Ut_0^4} \right] z^2
      + \cdots \Bigg\} \frac{\d U}{\Ut_0} ~,
   \label{tauorigseries}
 \end{eqnarray}
 where the value of $\Uth$ at $z = 0$ is
 \begin{equation}
   \Ut_0  = \sqrt{\frac{2 M_3}{U} + f_2 + \frac{\Lambda U^2}{3}}\; ~.   \label{Ut0}
 \end{equation}
 Taking the $z$ derivative of \er{tauorigseries} gives 
 \begin{eqnarray}
\fl
&&   \td{\tau}{z}  = \left\{ 1 - \left( \frac{M_4}{U} + \frac{f_3}{2} \right) \frac{z}{\Ut_0^2}
   - \left[ \left( \frac{M_5}{U} + \frac{f_4}{2} \right) \frac{1}{\Ut_0^2}
   - \left( \frac{M_4}{U} + \frac{f_3}{2} \right)^2 \frac{3}{2 \Ut_0^4} \right] z^2
   + \cdots \right\}_\wedge \td{\Uh}{z} \nn \\
\fl
&&    + \int_0^{\Uh} \left\{ - \left( \frac{M_4}{U} + \frac{f_3}{2} \right) \frac{1}{\Ut_0^2}
   - \left[ \left( \frac{M_5}{U} + \frac{f_4}{2} \right) \frac{1}{\Ut_0^2}
   - \left( \frac{M_4}{U} + \frac{f_3}{2} \right)^2 \frac{3}{2 \Ut_0^4} \right] z
   + \cdots \right\} \frac{\d U}{\Ut_0} ~.
   \label{tauzorigseries}
 \end{eqnarray}
 At $z = 0$, these become
 \begin{eqnarray}
   \left. \Uh \right|_0  & = & \left. \frac{\Rh}{z} \right|_0
      \approx \frac{\Rh_1 z}{z} = \Rh_1 ~, \\
   \left. \td{\Uh}{z} \right|_0 & =& \left. \frac{1}{z} \td{\Rh}{z} - \frac{\Uh}{z^2} \right|_0
      \approx \frac{(\Rh_1 + 2 \Rh_2 z)}{z} - \frac{(\Rh_1 z + \Rh_2 z^2)}{z^2}
      = \Rh_2 ~, \\
   \tau_0 & = & \int_0^{\Rh_1} \frac{dU}{\Ut_0}   \label{tau0} ~, \\
   \tau_1 & = & \Rh_2 - \int_0^{\Rh_1} \left( \frac{M_4}{U} + \frac{f_3}{2} \right) \frac{dU}{\Ut_0^3} ~.   \label{tau1}
 \end{eqnarray}

   From \er{Rhz} we see that
 \begin{equation}
   \rh_1  = \Rh_1 ~.
 \end{equation}
 We next substitute these series into \er{phiz1}, re-expand them as Taylor series, and solve for each power of $z$ in turn.  The $f_i$ then follow from \er{Wz}.  We find
 \begin{eqnarray}
\fl
   \rh_2 & =& \frac{\Rh_1}{2} + \Rh_2 ~,~~~~
      \rh_3 = \frac{2 \Rh_2}{3} + \Rh_3 + \frac{M_3}{2} ~,~~~~
      \rh_4 = \frac{\Rh_2 M_3}{2 \Rh_1} + \frac{3 \Rh_3}{4} + \Rh_4 + \frac{M_4}{3} + \frac{5 M_3}{8} ~,
      \label{rhi} \\
\fl
   f_2 & =& 1 - \frac{2 M_3}{\Rh_1} - \frac{\Lambda \Rh_1^2}{3} ~,~~~~
      f_3 = - 1 + (\Rh_1 + 2 \Rh_2) \left( \frac{M_3}{\Rh_1^2} - \frac{\Lambda \Rh_1}{3} \right) - \frac{2 M_4}{\Rh_1} ~.
      \label{fi}
 \end{eqnarray}
 Substituting for $f_2$ from \er{fi} into \er{tau0} with \er{Ut0}, the expression
 \begin{equation}
   \tau_0  = \int_0^{\Rh_1} \frac{dU}{\sqrt{\frac{2 M_3}{U} - \frac{2 M_3}{\Rh_1} + \frac{\Lambda U^2}{3}
   - \frac{\Lambda \Rh_1^2}{3} + 1}\;}
   \label{tau0O}
 \end{equation}
 allows one to iteratively solve for $M_3$ using known $\tau_0$ and $\Rh_1$ values.  Similarly, substituting for $f_3$ from \er{fi} into \er{tau1} allows $M_4$ to be found via another iterative process.  Knowing the $M_i$, one obtains the $f_i$ and $\rh_i$ from \er{fi} \& \er{rhi}.  Obviously, one gets $t_B$ from \er{tauf} 
 \begin{eqnarray}
   t_{B,0} = t_0 - \tau_0 ~,~~~~~~ 
   t_{B,i} = - (\rh_i + \tau_i) ~.
   \label{tBiO}
 \end{eqnarray}
 Thus we can evaluate all coefficients at a small distance away from the origin, so that the numerical integration can proceed.

 \subsection{Maximum}
 \label{max}
 
 A common feature of many big-bang cosmologies is that the diameter distance $\Rh(z)$ has a maximum $\Rh_m$ that occurs where our past null cone crosses the apparent horizon \cite{KraHel04b,Hel06,LH07,MH08,Hel09}.  Subscript $m$ will indicate evaluation at this maximum.  By definition, this locus has
 \begin{equation}
   \left. \td{\Rh}{z} \right|_m = 0 ~,
 \end{equation}
 and it follows from \er{dRdr} \& \er{vel} that
 \begin{equation}
   \Rth_m = W_m ~~~~~~\LRa~~~~~~ 1 - \frac{2 M_m}{\Rh_m} - \frac{\Lambda \Rh_m^2}{3} ~,
   \label{AHM}
 \end{equation}
 while \er{Wz}, \er{Rhrr} and \er{dmz} show
 \begin{equation}
   \left. \tdt{\Rh}{r} \right|_m  = - \left[\frac{M'}{W \Rh} \right]_m ~~~~~~\LRa~~~~~~
   \left. \tdt{\Rh}{z} \right|_m = - \left[ \frac{\varphi}{W \Rh} \td{M}{z} \right]_m ~,
   \label{RhzMax} 
 \end{equation}
 and
 \begin{equation}
   \left. \td{M}{z} \right|_m  = \Bigg\{  - \td{\tau}{z}
      - W B \Bigg[ \Bigg( \frac{M}{\Rh^2} - \frac{\Lambda \Rh}{3} \Bigg) \varphi
      + \frac{W}{(1 + z)} \Bigg] \Bigg\}_{\!\!m} \Bigg/ \Bigg[ A - \frac{B}{\Rh} \Bigg]_{\!m} ~.
      \label{MzMax}
 \end{equation}
 Now the $\varphi$ DE \er{phiz} has $\tdil{\Rh}{z}$ in the denominator, but \er{RhzMax} and \er{MzMax} ensure that the corresponding numerator,
 \begin{equation}
   \fl
   \left( \tdt{\Rh}{z} - \frac{\varphi}{\left( A \Rh - B \right)} \Bigg[ \frac{1}{W} \td{\tau}{z}
      + B \left\{ \left( \frac{M}{\Rh^2} - \frac{\Lambda \Rh}{3} \right) \varphi
      + \frac{W}{(1 + z)} \right\} \Bigg] \right)_m = 0 ~,
      \label{AHphi}
 \end{equation}
 is zero.  Similarly, \er{Wz} contains $0/0$, and it is this expression that is understood wherever $W$ appears in \er{dmz} and \er{phiz}.  So although $\tdil{\varphi}{z}$ and $\tdil{M}{z}$ are not actually divergent, their DEs are not suitable for numerical integration here.  Therefore we seek a Taylor series solution in the neighbourhood of $\Rh_m$, in powers of $\Delta z = z - z_m$,
 \begin{eqnarray}
\fl
   && \rh = \rh_m + \sum_{i=1}^\infty \rh_i \Delta z^i ~, \quad
      \Rh = \Rh_m + \sum_{i=2}^\infty \Rh_i \Delta z^i ~, \quad
      \tau = \tau_m + \sum_{i=1}^\infty \tau_i \Delta z^i ~, \quad \nn \\
\fl
   && M = M_m + \sum_{i=1}^\infty M_i \Delta z^i ~, \quad
      W = W_m + \sum_{i=1}^\infty W_i \Delta z^i ~, \quad
      f = f_m + \sum_{i=1}^\infty f_i \Delta z^i ~.
 \end{eqnarray}

   Again expressing the $\tau$ integrand as a Taylor series gives
 \begin{eqnarray}
   \tau & \approx & \int_0^{\Rh} \Bigg\{ 1 - \left( \frac{f_1}{2} + \frac{M_1}{R} \right) \frac{Z}{\Rt_m^2} \nn \\
   &&~~~~~~~~ - \left[ \Rt_m^2 \left( \frac{f_2}{2} + \frac{M_2}{R} \right)
      - \frac{3}{2} \left( \frac{f_1}{2} + \frac{M_1}{R} \right)^2 \right] \frac{Z^2}{\Rt_m^4} ~,
   + \cdots \Bigg\} \frac{\d R}{\Rt_m}
   \label{taumaxseries}
 \end{eqnarray}
 where
 \begin{equation}
   \Rt_m  = \sqrt{\frac{2 M_m}{R} + f_m + \frac{\Lambda R^2}{3}}\; ~,
   \label{Rtm}
 \end{equation}
 and thus its $z$ derivative is
 \begin{eqnarray}
\fl
   && \td{\tau}{z}  \approx
      \Bigg\{ 1 - \left( \frac{f_1}{2} + \frac{M_1}{\Rh} \right) \frac{Z}{\Rt_m^2}
      - \left[ \Rt_m^2 \left( \frac{f_2}{2} + \frac{M_2}{\Rh} \right)
      - \frac{3}{2} \left( \frac{f_1}{2} + \frac{M_1}{\Rh} \right)^2 \right] \frac{Z^2}{\Rt_m^4}
      + \cdots \Bigg\} \td{\Rh}{z} \nn \\
\fl
   &&~~~~~~~~ + \int_0^{\Rh} \Bigg\{ - \left( \frac{f_1}{2} + \frac{M_1}{R} \right) \frac{1}{\Rt_m^2} \nn \\
\fl
   &&~~~~~~~~~~~~~~~~~~ - \left[ \Rt_m^2 \left( \frac{f_2}{2} + \frac{M_2}{R} \right)
      - \frac{3}{2} \left( \frac{f_1}{2} + \frac{M_1}{R} \right)^2 \right] \frac{2 Z}{\Rt_m^4}
      + \cdots \Bigg\} \frac{dR}{\Rt_m} ~,
   \label{tauzmaxseries}
 \end{eqnarray}
 where $\tdil{\Rh}{z}|_m = 0$, so that the values at the maximum are
 \begin{eqnarray}
   \tau_m & = & \int_0^{\Rh_m} \frac{dR}{\Rt_m} ~,
      \label{taum} \\
   \tau_1 & = & - \int_0^{\Rh_m} \left( \frac{f_1}{2} + \frac{M_1}{R} \right) \frac{\d R}{\Rt_m^3} ~.
      \label{tau1m}
 \end{eqnarray}
 The series solution of the $\varphi$ DE, multiplied by $W \tdil{\Rh}{z}$, produces%
 \footnote{Note that \er{rh1} also satisfies \er{RhzMax} evaluated at $z_m$}
 \begin{eqnarray}
\fl
   \rh_1 & = & \frac{- 2 W_m \Rh_m \Rh_2}{M_1} ~,
      \label{rh1} \\
\fl
   \rh_2 & = & \left\{ \frac{f_1}{4 W_m M_1} - \left( \frac{1}{(1 + z_m)}
      + \frac{2 M_2}{M_1} \right) \frac{W_m}{2 M_1} \right\} \Rh_m \Rh_2 ~,
      \label{rh2} \\
\fl
   \rh_3 & = &
   - \frac{4 W_m \Rh_m \Rh_4}{3 M_1}
   + \left\{ \left( \frac{M_2}{M_1} - \frac{1}{2 (1 + z_m)} \right) W_{{m}}
   - \frac{f_1}{4 W_m} \right\} \frac{\Rh_m \Rh_3}{M_1}
   - \frac{2 W_m \Rh_2^2}{9 M_1} \nn \\
\fl
   &&~~~~ + \Bigg\{  \left(\frac{4 M_2}{3 M_1 (1 + z_m)}
   + \frac{1}{2 (1 + z_m)^2}
   + \frac{2 M_3}{M_1}
   - \frac{2 M_2^2}{M_1^2} \right) W_m \nn \\
\fl
   &&~~~~ - \left[  \left( \frac{1}{(1 + z_m)}
   - \frac{M_2}{M_1} \right) f_1
   + f_2 \right] \frac{1}{3 W_m}
   + \frac{f_1^2}{8 W_m^3} \Bigg\} \frac{\Rh_m \Rh_2}{3 M_1} ~,
   \label{rh3}
 \end{eqnarray}
 and feeding these into \er{Wz} times $\tdil{\Rh}{z}$ gives
 \begin{eqnarray}
   W_m & = & \sqrt{1 + f_m}\; ~, \\
   f_1 & = & \left( \frac{1}{(1 + z_m)} + (\Lambda \Rh_m^2 - 1) \frac{\Rh_2}{M_1} \right) 2 W_m^2 - \frac{2 M_1}{\Rh_m} ~, \nn \\
   f_2 & = & \Bigg\{ \frac{\left( 1 + \Rh_m^4 \Lambda^2 - 2 \Lambda \Rh_m^2 \right) 3 \Rh_2^2}{2 M_1^2}
      - \left( \frac{2}{\Rh_m} + \frac{(\Lambda \Rh_m^2 - 1) M_2}{M_1^2} \right) \Rh_2 \nn \\
   &&~~~~ + \frac{(\Lambda \Rh_m^2 - 1) 3 \Rh_3}{2 M_1}
      + \frac{(4 \Lambda \Rh_m^2 - 4) \Rh_2}{M_1 (1 + z_m)}
      + \frac{1}{(1 + z_m)^2} \Bigg\} W_m^2 \nn \\
   &&~~~~ -\frac{M_1}{\Rh_m (1 + z_m)}
      - \frac{2 M_2}{\Rh_m}
      - (\Rh_m^2 \Lambda - 1) \frac{3 \Rh_2}{2 \Rh_m} ~.
   \label{fim}
 \end{eqnarray}
 Now \er{AHM} gives $M_m$, and this plus \er{taum} with \er{Rtm},
 \begin{equation}
   \tau_m = \int_0^{\Rh_m} \frac{dR}{\sqrt{\frac{2 M_m}{R} + f_m + \frac{\Lambda R^2}{3}}\;} ~,
 \end{equation}
 allow $f_m$ to be found recursively.
 Similarly, putting $f_1$ from \er{fim} into \er{tau1m}
 \begin{equation}
   \tau_1  = - \int_0^{\Rh_m} \left[ \left( \frac{1}{(1 + z_m)} + (\Lambda \Rh_m^2 - 1) \frac{\Rh_2}{M_1} \right) W_m^2
   - \frac{M_1}{\Rh_m} + \frac{M_1}{R} \right] \frac{\d R}{\Rt_m^3} ~,
 \end{equation}
 permits a recursive solution for $M_1$, so that $f_1$ and $\rh_1$ follow from \er{fim} and \er{rh1}.  This applies to all higher orders, but the coefficient expressions rapidly become large.
 However, for $t_B$, \er{tauf} leads to 
 \begin{equation}
   t_{B,m}  = t_0 - (\rh_m + \tau_m) ~,~~~~~~
   t_{B,i} = - (\rh_i + \tau_i) ~,
 \end{equation}
 and $\rh_m$ is not determined above.  It must be fixed by making the maximum-series curve for $t_B(z)$ join up with the numerical curved for $t_B(z)$ a small $z$-distance before the maximum.%
 \footnote{With the distance-number counts-redshift method of earlier papers, $W_m$ and $M_1$ could only be determined by matching the maximum series to the numerical integration at some point just before the maximum.  The difference is probably because a differential equation for $\tdil{M}{z}$  in that method has been replaced by an integral equation for $\tau$ here.  However, each $M$ coefficient now requires an iterated integration, which was not necessary previously.}

   These properties of the maximum in $\Rh$ are not merely difficulties to be overcome, they provide independent information with which the results of numerical integrations may be cross-checked \cite{Hel06,LH07,MH08}.  (In fact \er{AHM} applies also in the case of non-zero pressure \cite{AlfHel10}.)  If one can measure $\Rh_m$, then, for given $\Lambda$, \er{AHM} immediately gives the total gravitational mass within this radius.  This holds regardless of the magnitude of any intervening radial inhomogeneity, and it holds only at $\Rh_m$.  As we will see, this $M$ must agree with the $M$ (and $W$) obtained from numerical integration.

 \section{Data Functions}
 \label{data}

The algorithm outlined above
requires us to provide $\tau$, $\Rh$, $\td{\tau}{z}$, $\tdtil{\Rh}{z}$, etc at any given $z$.
Real observations merely give us a set of discrete values. Hence there
is a problem of
 how to calculate higher derivatives based on discrete values.
In the case of ideal measurements one could use numerical algorithms
to estimate the derivatives, 
 for example the three-point or
 five-point derivation rules \cite{NumRec}. 
In practice however, the data are too noisy (supernova data at $z\approx 0.4 - 1$)
or there are too few measurements (age data, and supernova data for $z > 1$).

One method for extracting a smooth curve from the data would be to apply a procedure like the moving average. Such a procedure was outlined in Ref. \cite{MH08}. 
Another would be to apply some kind of a smoothing, like 
 Gaussian smoothing
-- for details see \cite{ede7}.
The question is how to smooth the data without distorting it,
or in other words how to distinguish between 
 noise and real variability of the data.
In the literature this problem is mostly investigated  in the context
of a reconstruction the evolution of dark energy
where the data also needs to be differentiated twice
to constrain the dark energy equation of state
 \cite{ede1,ede2,ede3,ede4,ede5,ede6,ede7,ede8,ede9}.

\begin{figure}
 \begin{center}
 \includegraphics[scale=0.4]{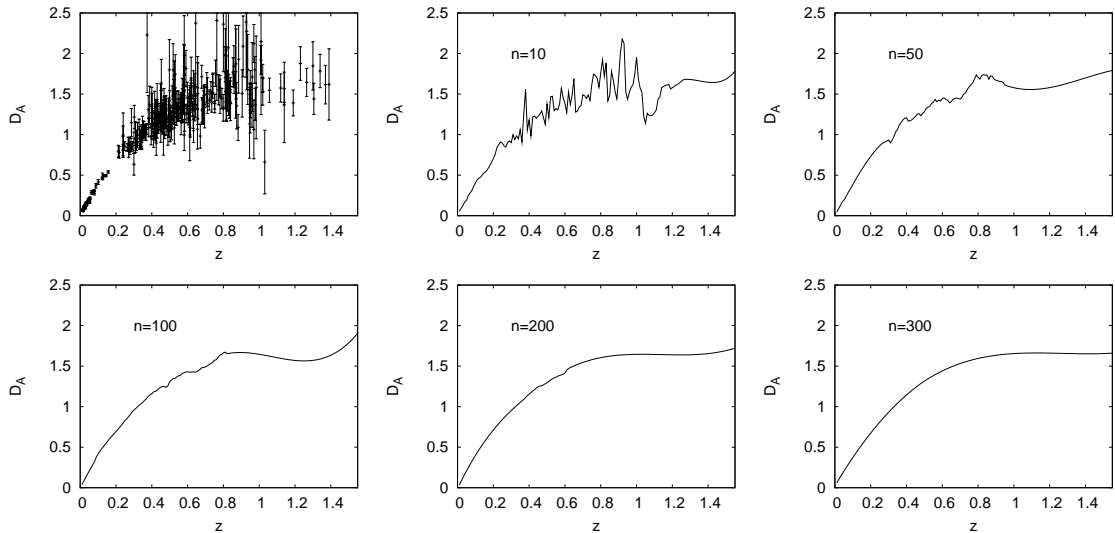}
 \caption{ Supernova data (upper left panel) and the different fits (other panels).  For each graph, the value of $n$ denotes the bin size: within each redshift interval that contains $n$ supernovae, a 3rd order polynomial is fitted to the data.
 \label{ffn}
 }
 \end{center}
 \end{figure}

To highlight the problem, let us for a moment focus on one 
procedure of smoothing the data.  
 Let us consider the supernova data, $D_L(z)$, in a given redshift bin, and apply the least squares method to fit a polynomial to the points in the bin. Now, 
 the bin is moving, i.e.
for each supernova we consider $n-1$ supernovae around it and
we fit a polynomial to these $n$ points. Then we move to the next supernova 
and repeat the procedure\footnote{The size of the interval will thus vary with redshift.  Alternatively one can fix the redshift interval, but then the number of points in each interval will no longer be constant.}.
On the one hand, the smaller $n$ is, the smaller the redshift interval, and therefore the more the flexibility of a fit --- one can extract more detailed information about the curve.  
On the other hand, if $n$ is too small then the fitting procedure picks up the `noise' of the data.  This is illustrated in Fig.\ \ref{ffn}.  If the number of supernovae within an interval is $n=10$, then the noise becomes significant.  Even with $n=200$ we can still see some artifacts like the minimum around $z=1.3$, which is rather unphysical.  This shows that we still need a much larger large number of observations to generate a reliable average $D_L(z)$ curve.%
 \footnote{To investigate angular variation, or equivalently check isotropy, we would need far more data in {\it each} directional bin.}  
 At this point however, when we only have a few hundreds of supernova measurements, and a mere 30-odd age estimations.
The present data is simply not good enough to be used in the algorithm outlined above.
The situation will obviously improve over the coming years.

Still, we would like to use the real data to see what kind of results 
can be obtained when applying our algorithm.
We therefore decided to apply a crude approach and  to fit a chosen curve to the whole data.   Of course,
  this is something that one would prefer to avoid.
The idea of the inverse problem is to use pure data,
without any assumptions of its form and use it to specify the geometry of the spacetime.
Nevertheless, let us proceed in this way, bearing in mind
that once the data is of better quality and large 
 quantity, it will be possible
to apply the method in its true spirit.

 Inevitably, the choice of a fitting function must affect the form of the output.
In order to minimise this bias 
 we consider several different functions, so that we 
 can assess the impact of such choices on the results of our analyses.

\begin{figure}
 \begin{center}
 \includegraphics[scale=0.6]{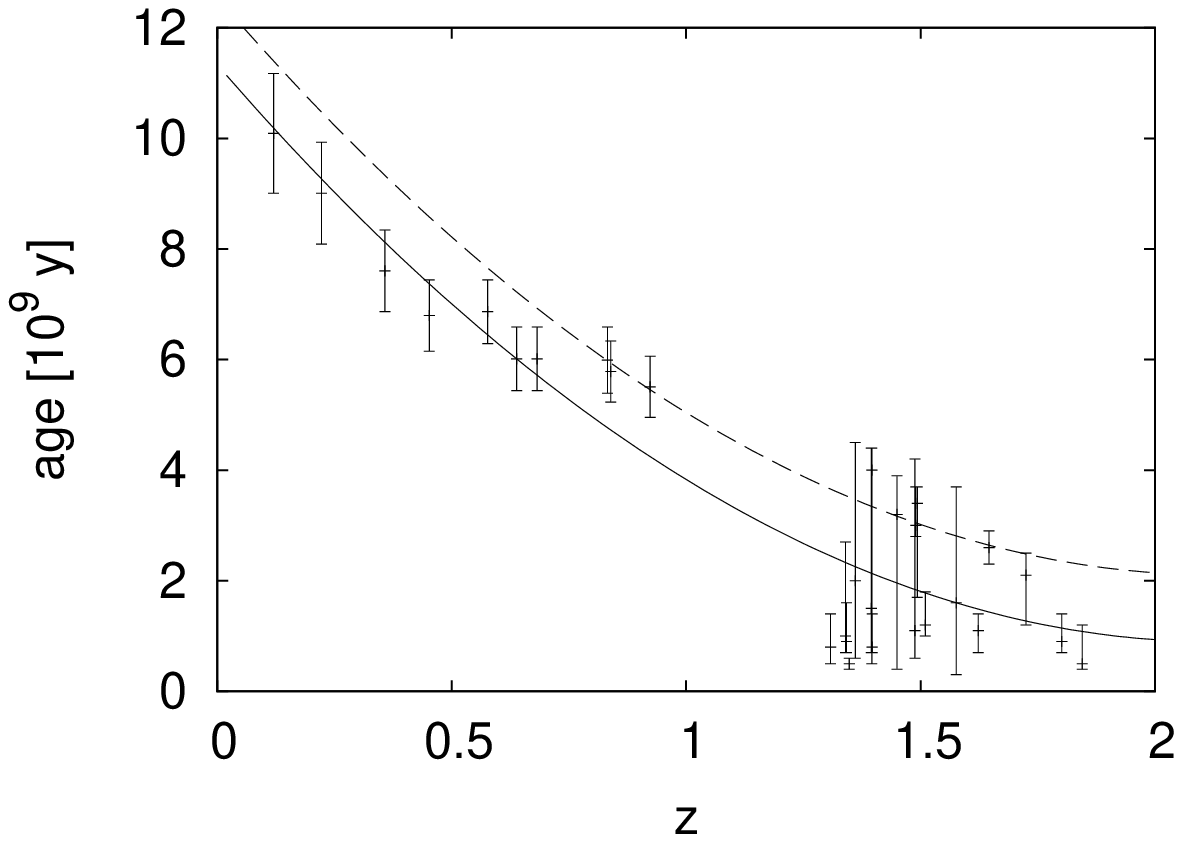}
 \includegraphics[scale=0.6]{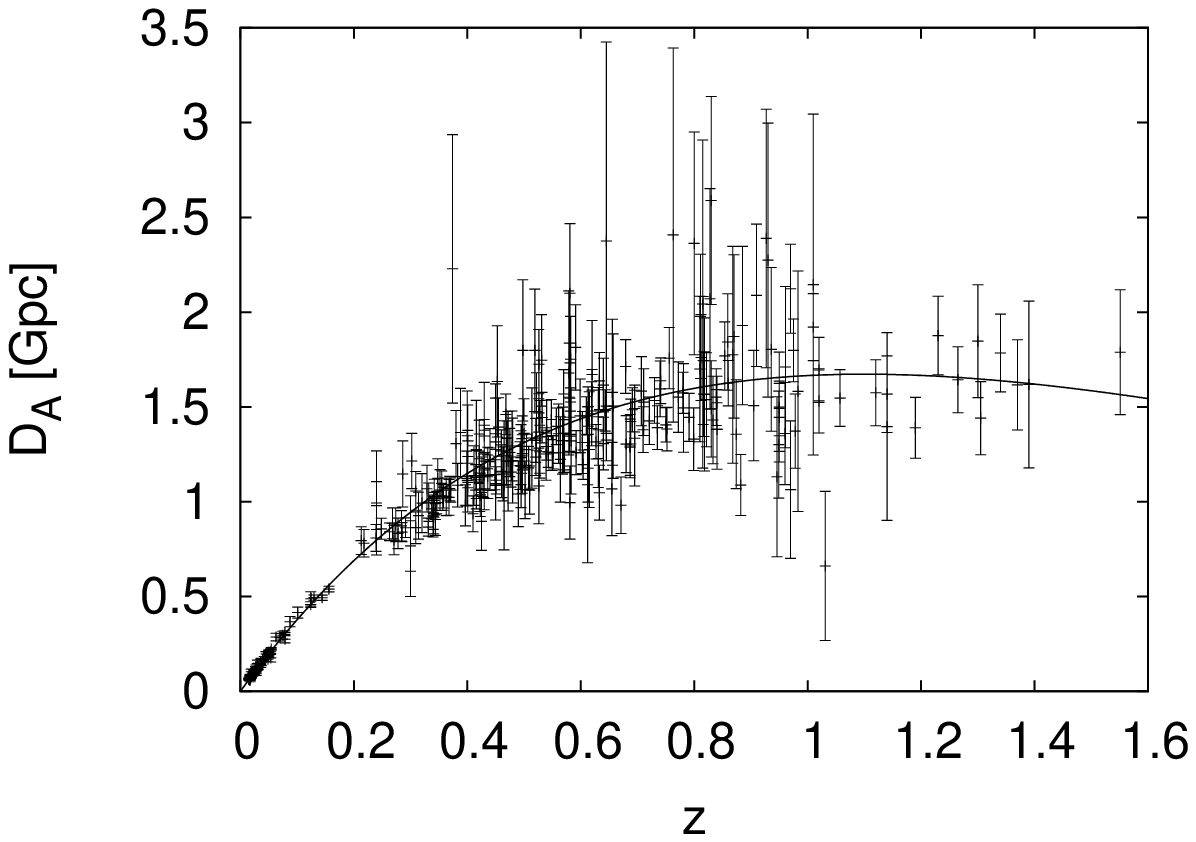}
 \caption{ {\em Left}: The maximum stellar age data, and the best-fit polynomial (quadratic) of eq.\ \er{tdf}, one of 3 different fitting functions tried.  Allowing for star formation time, the dashed line is the best fit + 1.21 Gy.
 {\em Right}: Angular diameter distance from supernova data, the solid line represents the best-fit relation \er{daf}, the first of 5 functions tried.
 \label{figdata}
 }
 \end{center}
 \end{figure}

 \begin{itemize}

 \item {\bf Age data} \\   
 The  maximum stellar 
age data we used comes from \cite{McC04,SVJ05,SJVKS09}, 
 and we considered 3 fitting functions:

 \begin{enumerate}

 \item Type 1
 \begin{equation}
   \tau(z) =  t_0 + t_1 z + t_2 z^2 ~,
   \label{tdf}
 \end{equation}
 where $t_0 = (11.3355 \pm 0.7804) \times 10^9$ y,
 $t_1= (-9.80635  \pm 1.559) \times 10^9$ y, and 
 $t_2= (2.3029 \pm 0.711) \times 10^9$ y (see Fig.\ \ref{figdata}).  

 \item Type 2
 \begin{equation}
   \tau(z) =  t_0 {\rm e}^{-t_1 z} ~,
 \end{equation}
 where $t_0 = (12.3137 \pm 0.9151) \times 10^9$ y and
 $t_1 = 1.21623 \pm 0.1046$.

 \item Type 3
 \begin{equation}
   \tau(z) =  \frac{a}{(1+z)^b} ~,
 \end{equation}
 where
 $a = (14.227  \pm 1.318) \times 10^9$ y, and 
 $b = 2.05036 \pm  0.1873$.  
 The above form is suggested by the Einstein-de Sitter time behaviour, where $\tau = (2/3) H_0 (1+z)^{-3/2}$.

 \end{enumerate}

 \item  {\bf Angular diameter distance data} \\

 The luminosity data, usually given in the form of distance moduli, $\mu$, where 
 \[ \mu = 5 \log \frac{D_A}{(1+z)^2} + 25 ~, \] 
 come from the supernova observations, here the Union data set \cite{K08}.  For our purpose, it is more convenient to express the functions in the form of the angular diameter distance $D_A(z)$.  We consider 5 different forms of fitting function:

 \begin{enumerate}

 \item Type 1
 \begin{equation}
   D_A(z) = z D_1 {\rm e}^{-D_2 z} ~,
   \label{daf}
 \end{equation}
 where               
 $D_1 = 4.13529  \pm 0.1195$ Gpc and 
 $D_2 = 0.909403  \pm 0.03956$ (see Fig.\ \ref{figdata}). 
 The Hubble constant for this fit is

 \[ H_0 =  \left(\left.\frac{{\rm d} D_A}{{\rm d} z}\right|_0\right)^{-1} = 72.496 \pm 2.095 {\rm ~km~s}^{-1}{\rm~Mpc}^{-1}~. \]

 \item Type 2
 \begin{equation}
   D_A(z) = D_1 z + D_2 z^2 + D_3 z^3 ~,
 \end{equation}
 where               
 $D_1 = 4.04282  \pm 0.1534$ Gpc,
 $D_2 = -3.2535  \pm 0.3655$ Gpc, and 
 $D_2 = 0.863957  \pm 0.2028$ Gpc.
 The Hubble constant for this fit is $H_0 = 74.154 \pm 2.814 {\rm ~km~s}^{-1}{\rm~Mpc}^{-1}.$

 \item Type 3
 \begin{equation}
   {D}_A = \frac{A}{1+z} {\rm atan} \left( z B \right) ~, 
 \end{equation}
 where               
 $A = 4.18101  \pm 0.293$ Gpc,
 $B= -1.0185  \pm 0.0913$.
 The Hubble constant for this fit is $H_0 = 70.401 \pm 8.822 {\rm ~km~s}^{-1}{\rm~Mpc}^{-1}.$

 \item Type 4
 \begin{equation}
   {D}_A = A \times {\rm atan} \left( zB \right) ~, 
 \end{equation}
 where               
 $A = 1.34115 \pm 0.04135$ Gpc and 
 $B= 2.95013  \pm 0.2161$.
 The Hubble constant for this fit is $H_0 = 75.771 \pm 6.022 {\rm ~km~s}^{-1}{\rm~Mpc}^{-1}$.  Note that the function ${\rm atan} (x)$ has a maximum at infinity, so no correction for systematics (see below) can be applied for these 2 choices.  As long as the data do not approach $\Rh_m$, there should be no problem.

 \item Type 5
 \begin{equation}
   {D}_A = D_{\Lambda CDM} ~,
 \end{equation}
 where $D_{\Lambda CDM}$ is the angular diameter distance as in the standard cosmological model -- the isotropic and homogeneous \LCDM\ model.  We use this case to check how this choice (which is the standard choice) influences the analysis.

 \end{enumerate}

 \end{itemize}

 \section{Correction for systematics}\label{corsys}

   It was shown in \cite{Hel06,MH08} that the functions from the maximum series solution will not match properly with the nearby numerical solutions, if the data contains any systematic error.  In other words, the data as measured does not correspond to a fully self-consistent metric solution, in the sense that the various relations that must hold at $\Rh_m$ will not be exactly true at $z_m$, or they will not all hold at quite the same $z$.  This will be the case with all real data.  
It was further demonstrated how these area-distance-maximum relations can be used to estimate a ``correction'' to the data so that a consistent solution is obtained.  The same consistency requirement arises for the present datasets and solution algorithm; they are the second of \er{AHM} and \er{AHphi}.

\begin{figure}
 \begin{center}
 \includegraphics[scale=0.6]{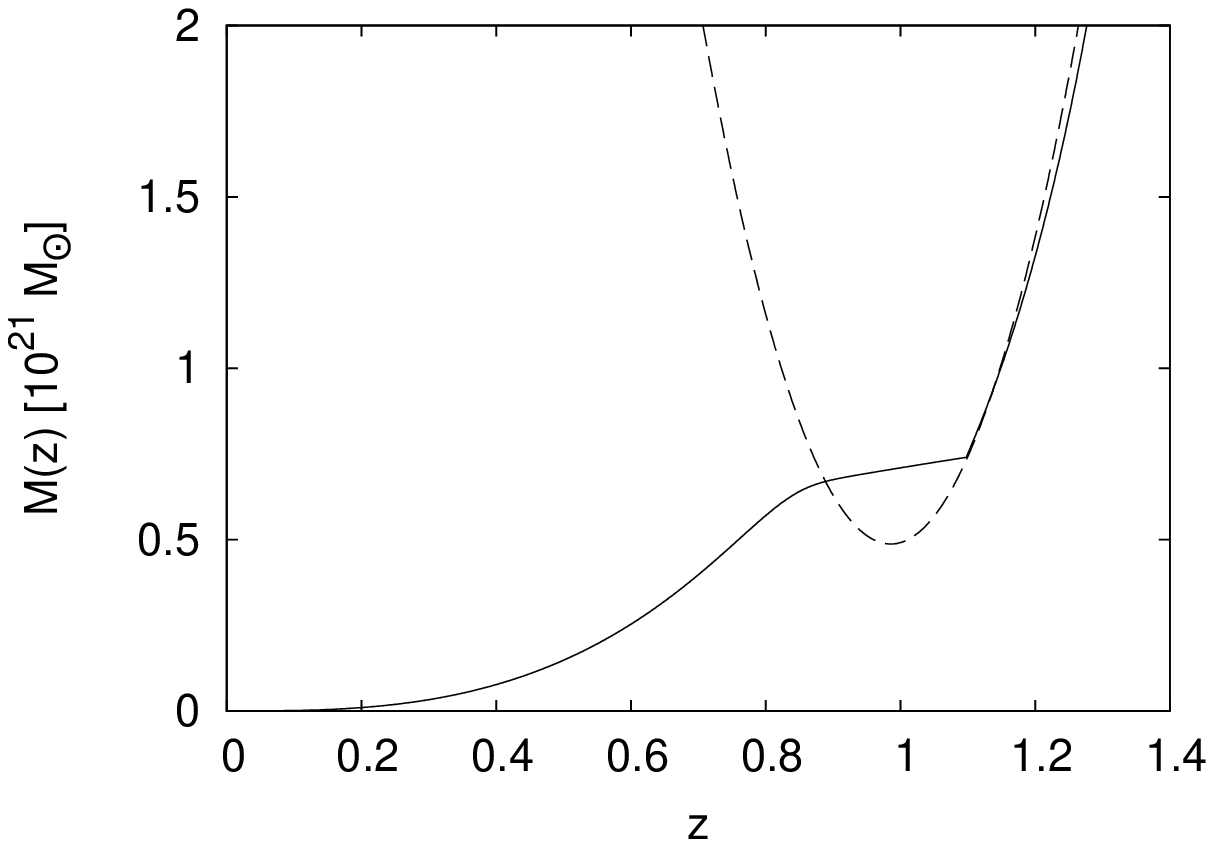}
 \includegraphics[scale=0.6]{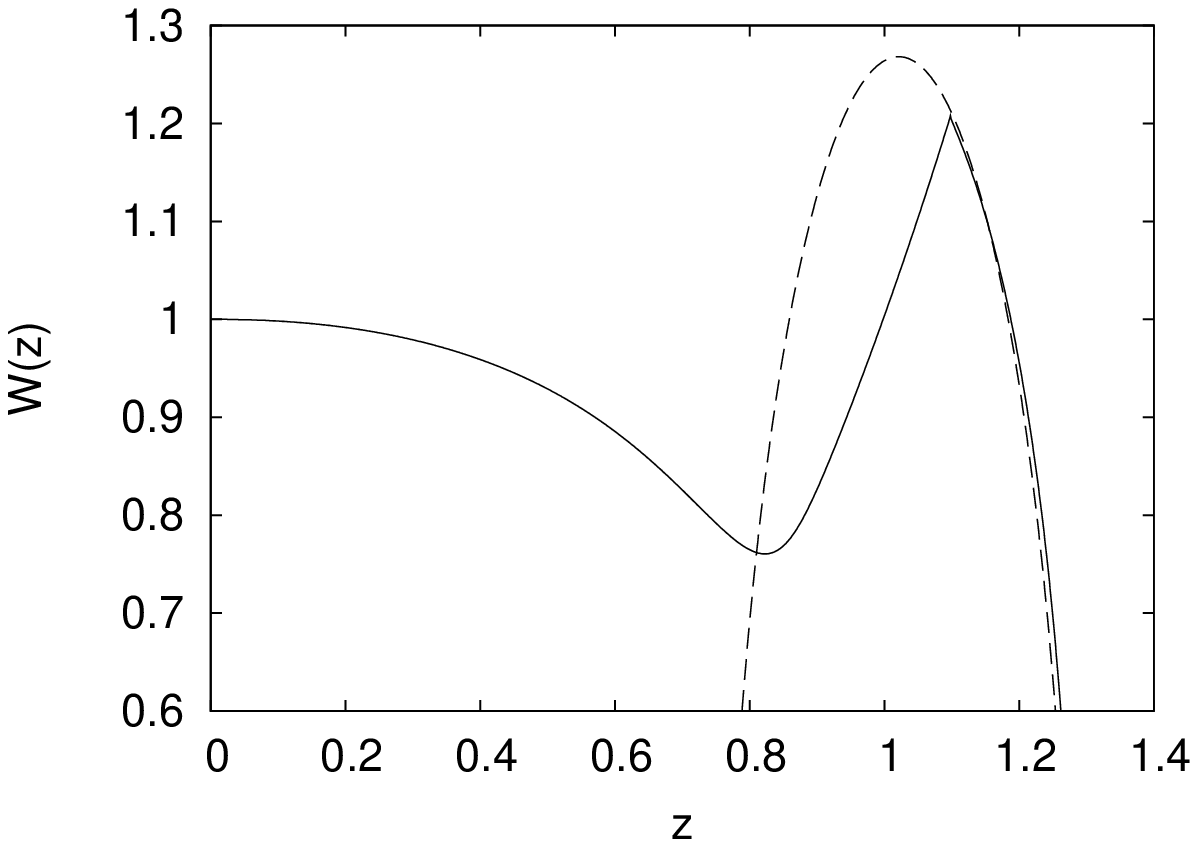}
 \caption{
 Results obtained with uncorrected data, for a model with  $\Omega_\Lambda = 0.7$.  The dashed line is the series solution near the maximum $\Rh_m$, and the solid line is the numerical solution.  It is evident that the series and numerical solutions do not match smoothly.  
 {\em Left}: $M(z)$.  {\em Right}: $W(z)$.
 \label{figbefcor}
 }
 \end{center}
 \end{figure}

   If we solve the equations without correction, then these relations will not be exactly satisfied, and we will encounter an undesirable numerical behaviour near the maximum --- see Fig.\ \ref{figbefcor}.  The point is that at $\Rh_m$ several relations must hold, and imperfect data will not satisfy them.  The observational data is always accompanied by systematics, therefore a correction is needed if a valid solution is to be obtained.  Note that this check detects the presence of a cumulative systematic error at $\Rh_m$, and enables one to find a correction that makes the data self-consistent.  However, the $\Rh_m$ conditions do not show what correction is needed {\it away from the maximum}, so an appropriate smoothness assumption must be made.

   Since the age data is probably the most sensitive to systematics.  For this paper, we boldly assume that the age of the oldest stars at each radius tracks the local age of the Universe.  In fact, the oldest observed stars, most probably, are not even of Population III, which are believed to be very massive and rapidly evolving.  Therefore we focus on ``correcting" the age data, and we propose the simplest correction, i.e. adding a constant to the age data, which corresponds to the fact that the oldest stars observed in galaxies formed some time after the Big Bang. 

   To demonstrate how the correction procedure is working, let us consider a model with $\tau(z)$ and $D_A(z)$ data of type 1, i.e. as given by \er{tdf} and \er{daf} respectively, and a cosmological constant, $\Lambda$, corresponding to $\Omega_\Lambda = 0.7$.

\begin{figure}
 \begin{center}
 \includegraphics[scale=0.6]{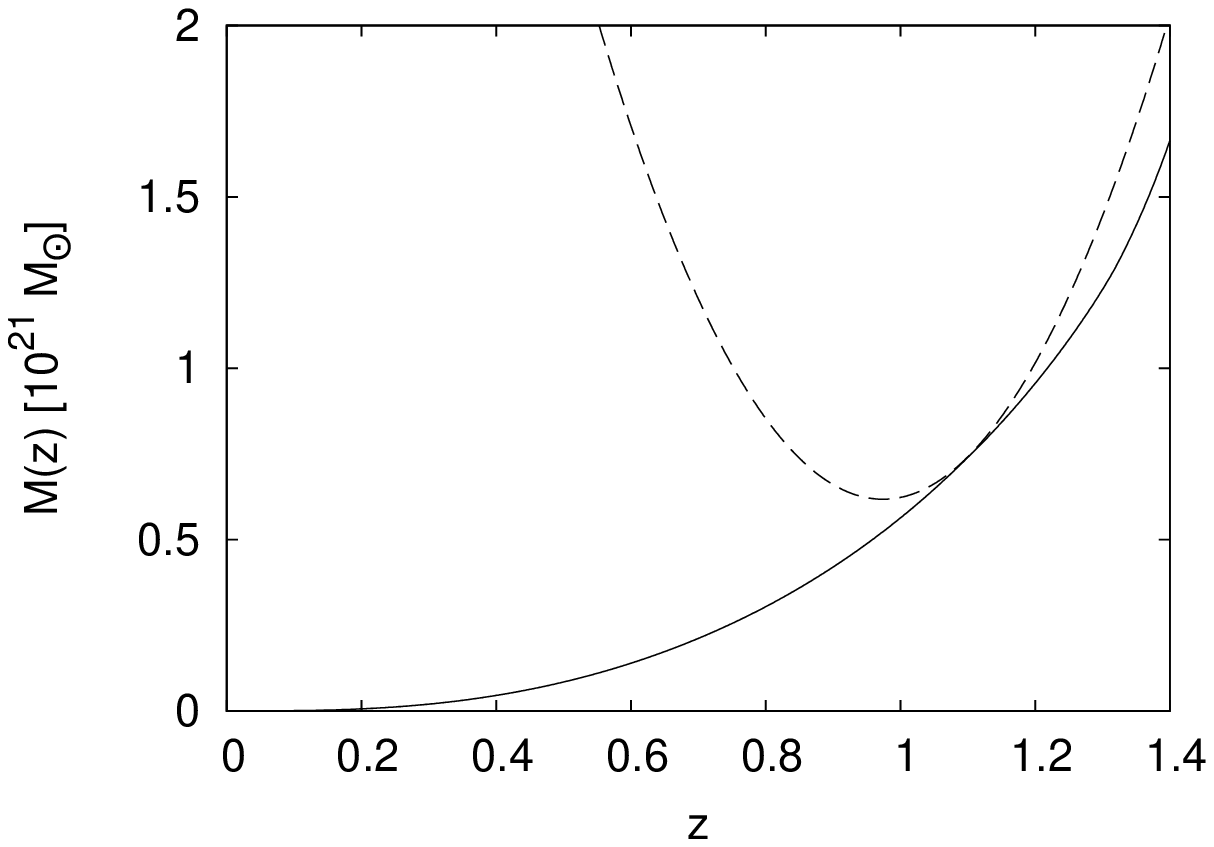}
  \includegraphics[scale=0.6]{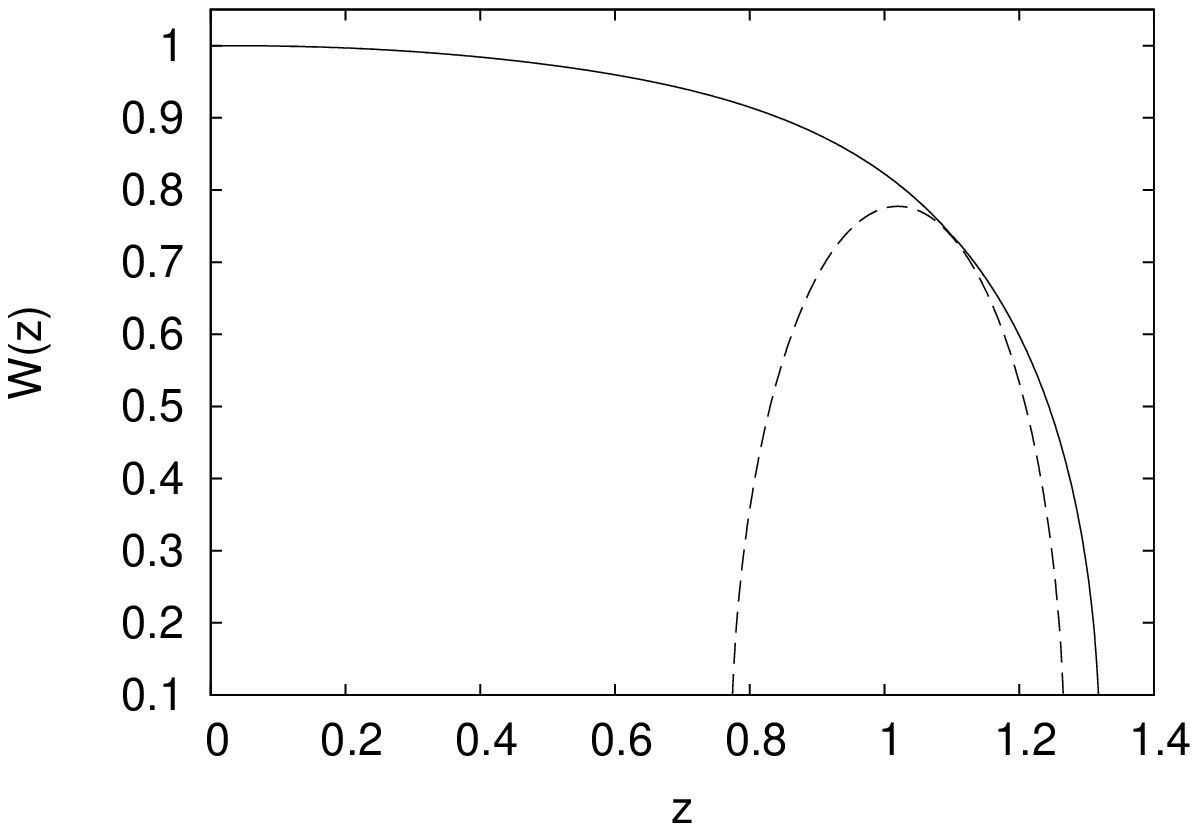}
  \caption{ Results after correcting the age data, showing the numerical integration (solid line) together with the series solution near $\Rh_m$ (dashed line) for the $\Omega_\Lambda = 0.7$ case, where the  correction to the age data is of form $t_0 \rightarrow t_0 + \Delta\tau$, where $\Delta\tau = 1.21 \times 10^9$ y.
 {\em Left}: $M(z)$.  {\em Right}: $W(z)$.  
 \label{figaftcor}
 }
 \end{center}
 \end{figure}

   Fig.\ \ref{figbefcor} presents the results obtained without our correction procedure.  The Taylor expansion solution in the vicinity of the maximum is plotted using a dashed line, the numerical integration with a solid line.  As seen, the two results do not match smoothly.  However, correcting the age data by adding a constant, i.e. $\tau \rightarrow \tau + \delta\tau$, results in a smooth fit.  The functions $M(z)$ and $W(z)$ after the correction are presented in Fig.\ \ref{figaftcor}.  The correction ($\delta\tau$) turns out to depend on the value of $\Lambda$.  For observational data in form of \er{tdf} and \er{daf} with $\Omega_\Lambda = 0.7$ (where $\Omega_\Lambda = (c^2 \Lambda)/(3 H_0^2)$), we find $\delta\tau = 1.21 \times 10^9$~y; for $\Omega_\Lambda = 0.4$ we find $\delta\tau = 0.94 \times 10^9$~y, and for $\Omega_\Lambda = 0$, $\delta\tau = 0.65 \times 10^9$~y.  Given the scatter in the data, and likely nature of the observed oldest stars, these values are surprisingly reasonable.

 \section{Algorithm}
 \label{alg}

   The above was turned into a numerical procedure, written in FORTRAN.  The following gives the key elements of the algorithm used.  
 \begin{enumerate}
 \item   \label{AlgO}   In the neighbourhood of the origin, fit low order polynomials to $D_A(z)$ and $\tau(z)$, and solve for the Taylor coefficients in \er{rhi}-\er{tBiO}, using a bisection method, as described at the end of section \ref{orig}.  These provide initial conditions at a small $z$ value, for the numerical integration of the DEs, which can't be evaluated at $z = 0$.  
 \item   Choose functional forms for $D_L(z)$ and $\tau(z)$, and do least-squares fits to the observational data.
 \item   \label{AlgM}  
At each step (each $z_i$) of the numerical integration, solve \er{Mz} to find the value of $M$ for the next step, $M_{i+1}$.
 \item   \label{AlgW}  Use $M_{i+1}$ and  \er{tbf} to find $W_{i+1}$.
 \item   \label{Algr}  Solve \er{phi} to find $r_{i+1}$.
 \item   \label{AlgtB}   Obtain $t_{B\,i}$ from \er{tauf}.
 \item   In practice steps \ref{AlgM}-\ref{AlgtB} can be combined into one integration routine, calculating all function values at each $z$ step.  We did this by means of a 4-th order Runge-Kutta method.
 \item   In the neighbourhood of the maximum, $\Rh_m$, do a low order polynomial fit to $D_L(z)$ and $\tau(z)$, and then calculate the coefficients of the near-maximum Taylor expansion given in section \ref{max}.  
 \item   \label{AlgC1}   Use the apparent horizon relations to perform an error correction, as explained in section \ref{corsys}.  
 \item   \label{AlgC2}   Once the data is corrected, start the procedure from the beginning, with the adjusted data as input.
 \item   In practice steps \ref{AlgC1}-\ref{AlgC2} need to be repeated until the outcome of the algorithm and the series expansion match smoothly --- see Sec.\ \ref{corsys} for details.  
 \end{enumerate}
 With $M(r)$, $E(r)$, and $t_B(r)$ in hand, the LT model is determined, and one can calculate its state and evolution for any instant using \er{tbf}.

 \section{Results}\label{res}

It should be re-iterated that our present purpose is to validate various possible methods of reducing observational data to metric information, and investigate how sensitive they are to variations in approach.  We do not suggest the LT models derived here represent reality.  Rather we are working towards a reliable method that can be used with much larger and more complete survey datasets.  
The ultimate goal is to have a working algorithm that allows the metric of the cosmos to be derived from the observations, rather than choosing a metric first and then trying to fit its functions to observations.

\begin{figure}
 \begin{center}
 \includegraphics[scale=0.7]{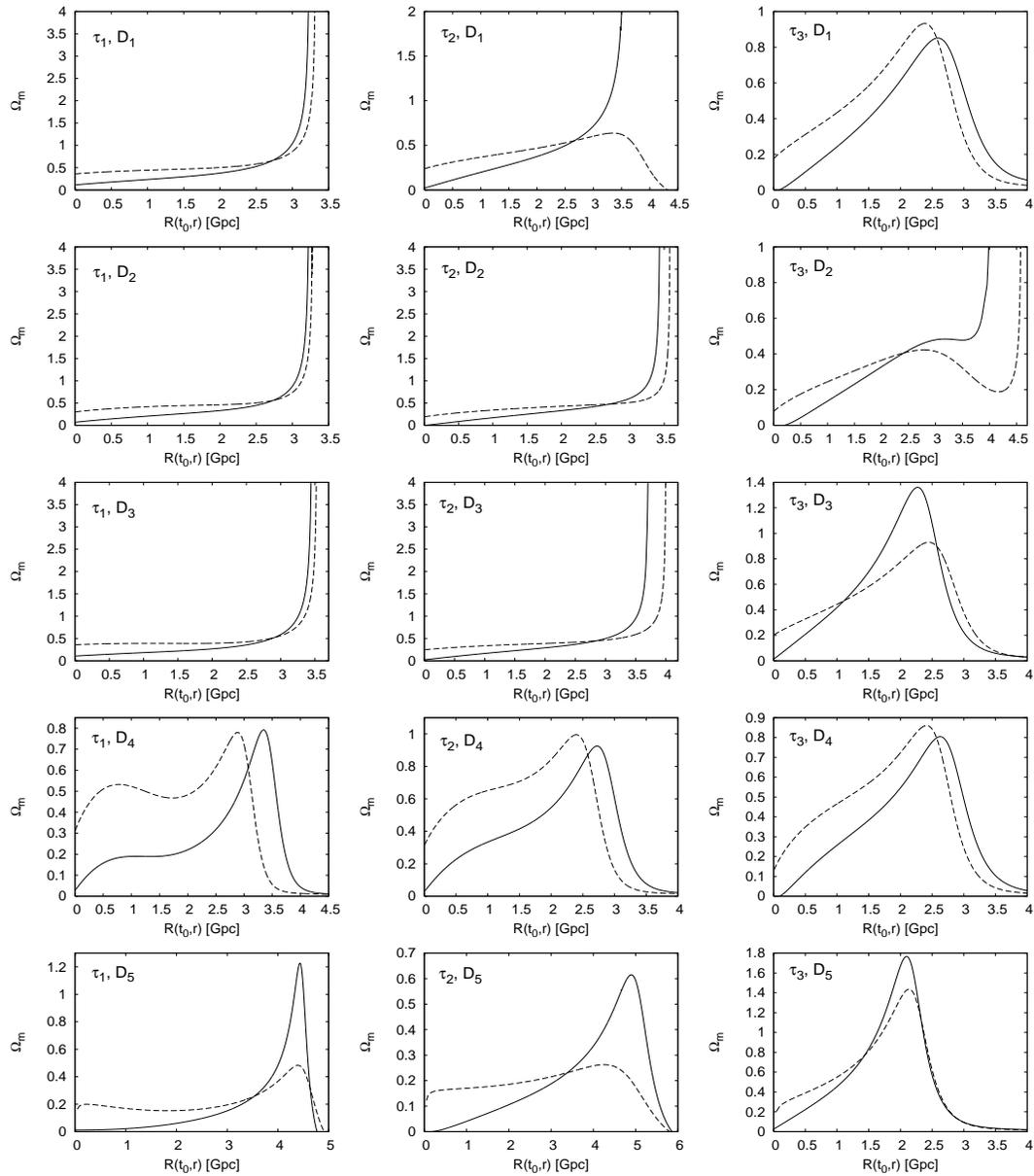}
 \caption{ Present-day 
density profile $\Omega_m(r)$ for models with fits to observations in the form of $\tau_i(z)$ and $D_j(z)$ (where $i$ and $j$ correspond to a fitting type --- see Sec.\ \ref{data}).  The pair of $\tau$ and $D$ is given in the upper left corner of each panel.  Solid line corresponds to a model with $\Omega_\Lambda = 0$ and dashed line to $\Omega_\Lambda = 0.7$.  (Note that these profiles do not represent observations, which are necessarily on the PNC.)  
 \label{rOm1} }
 \end{center}
 \end{figure}

\begin{figure}
 \begin{center}
 \includegraphics[scale=0.7]{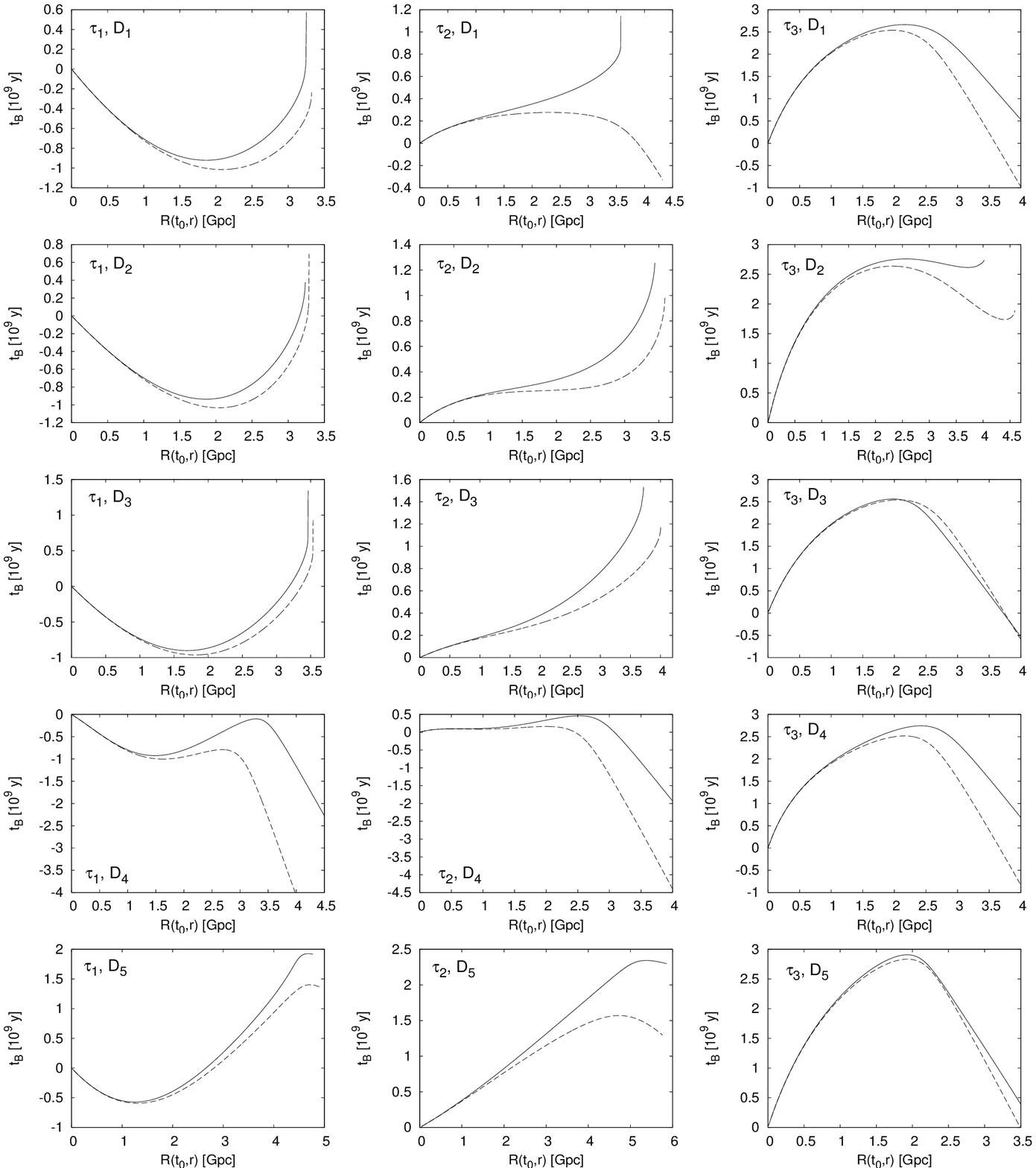}
 \caption{Bang time function $t_B(r)$ for models with fits to observations in the form of $\tau_i(z)$ and $D_j(z)$ (where $i$ and $j$ correspond to a fitting type --- see Sec.\ \ref{data}).  The pair of $\tau$ and $D$ is given in upper or lower left corner of each panel.  Solid line corresponds to a model with $\Omega_\Lambda = 0$ and dashed line to $\Omega_\Lambda = 0.7$.
 \label{rtb1} }
 \end{center}
 \end{figure}

   Let us now test different combinations of fitting functions for $\tau_i$ \& $D_j$ (where $i$ and $j$ correspond to a fitting type --- see Sec.\ \ref{data}).  We will consider both models with and without the cosmological constant, so all together we have 30 different models.  The results in form of $\Omega_r(r)$ and $t_B(r)$ are presented in Figs.\ \ref{rOm1} and \ref{rtb1} respectively.  The results for models with $\Lambda = 0$ are presented using solid lines, results for models with the standard model cosmological constant value ($\Omega_\Lambda=0.7$) are presented using dashed lines.  As may be seen, different types of algebraic fits lead to different results (scale and range also vary from panel to panel).  Clearly, not every pair of $\tau(z)$ \& $D_A(z)$ fit-functions produces a satisfactory model.  Some exhibit shell crossing at the current instant (though not along the past null cone):
$(\tau_1$\,\&\,$D_1)$, 
$(\tau_1$\,\&\,$D_2)$, 
$(\tau_2$\,\&\,$D_2)$, 
$(\tau_2$\,\&\,$D_3)$, and
$(\tau_3$\,\&\,$D_2)$.  
Others have $M'<0$ for low z which leads to negative density:
$(\tau_2$\,\&\,$D_5)$, 
$(\tau_3$\,\&\,$D_1)$,
$(\tau_3$\,\&\,$D_2)$, and
$(\tau_3$\,\&\,$D_4)$.
While others suffer a similar problem at high redshifts: 
$(\tau_1$\,\&\,$D_5)$, 
$(\tau_2$\,\&\,$D_1)$, and
$(\tau_2$\,\&\,$D_5)$.

 \subsection{Error estimation}
 \label{Ee}

   There are two sources of error in our analysis: the observational errors (both random and systematic), and the (systematic) errors introduced by the choice of function used to fit the data.  To estimate the `confidence region', or rather `error band',  we plotted all models that do not exhibit a shell crossing, or have $M'<0$.  In addition we also plotted results for models with $\tau \pm \Delta \tau$ and $D_A \pm \Delta D_A$, where $\Delta \tau$ and $\Delta D_A$ are uncertainties calculated separately for each fit
 \footnote{The error of the quantity $Q$, that depends on parameters $q_i$, i.e.  $Q = Q(q_i)$, is estimated using the propagation of uncertainty method: $( \Delta Q)^2 = \sum_i ({\partial Q}/\partial q_i)^2 (\Delta q_i)^2.$}.

 \subsection{Discussion}
 \label{D}

\begin{figure}
 \begin{center}
 \includegraphics[scale=0.5]{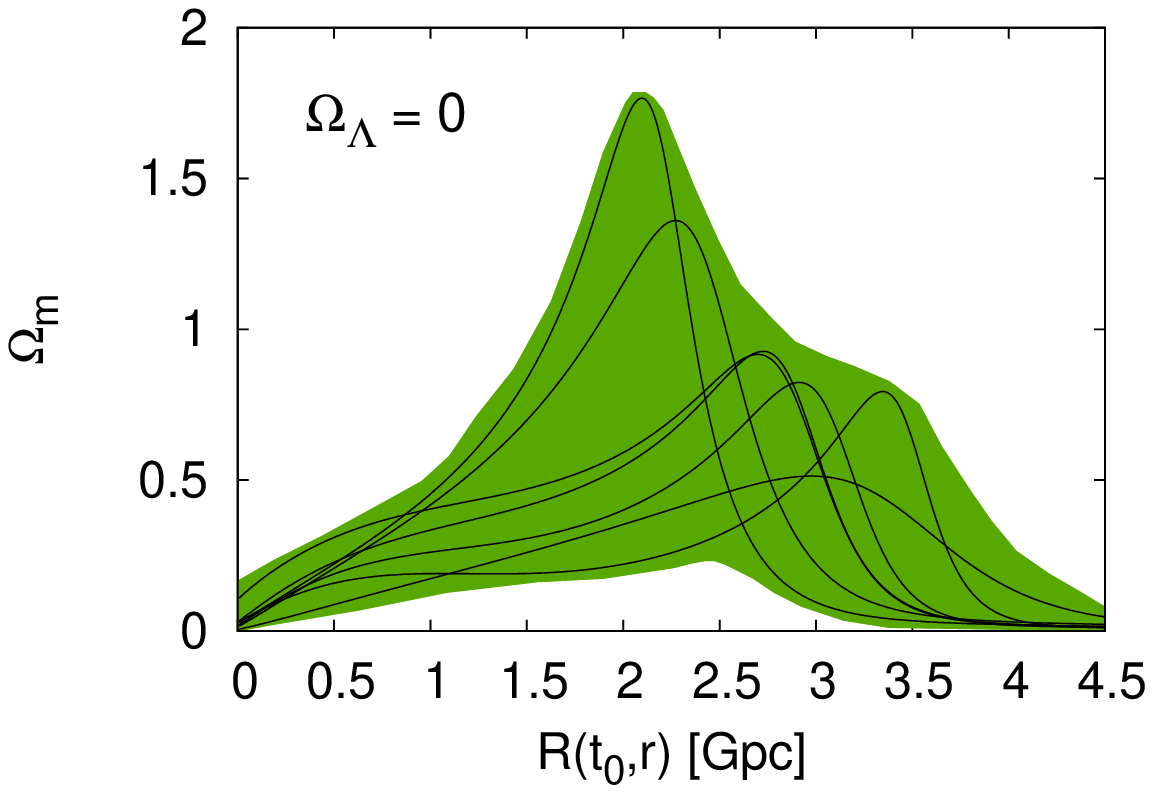}
 \includegraphics[scale=0.5]{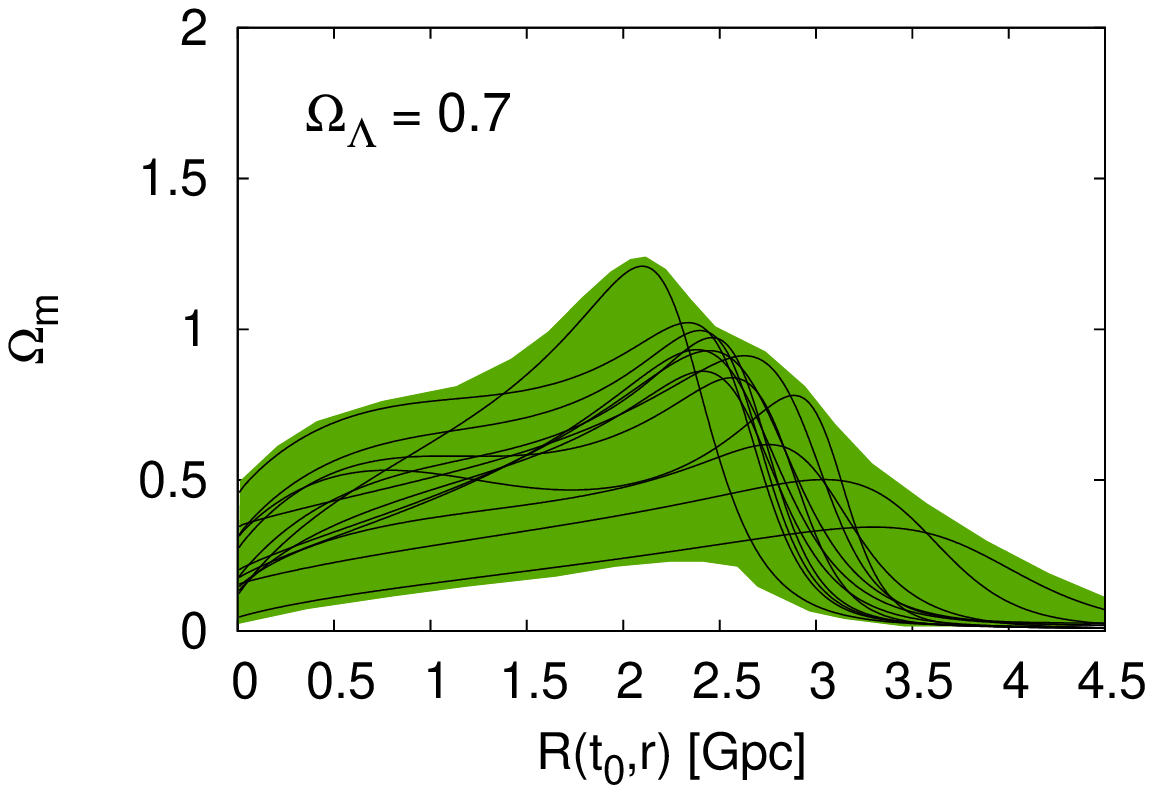}
 \caption{
 Constraints on the present-day density profile from luminosity distance and maximum stellar age data.  
 The curves are the individual fits under different assumptions.  The green region is the envelope of all fitted models satisfying some minimal reasonableness requirements (see text).  The region is large because there is still a lot of scatter and uncertainty in the data. \label{fge1} }
 \end{center}
 \end{figure}

\begin{figure}
 \begin{center}
 \includegraphics[scale=0.5]{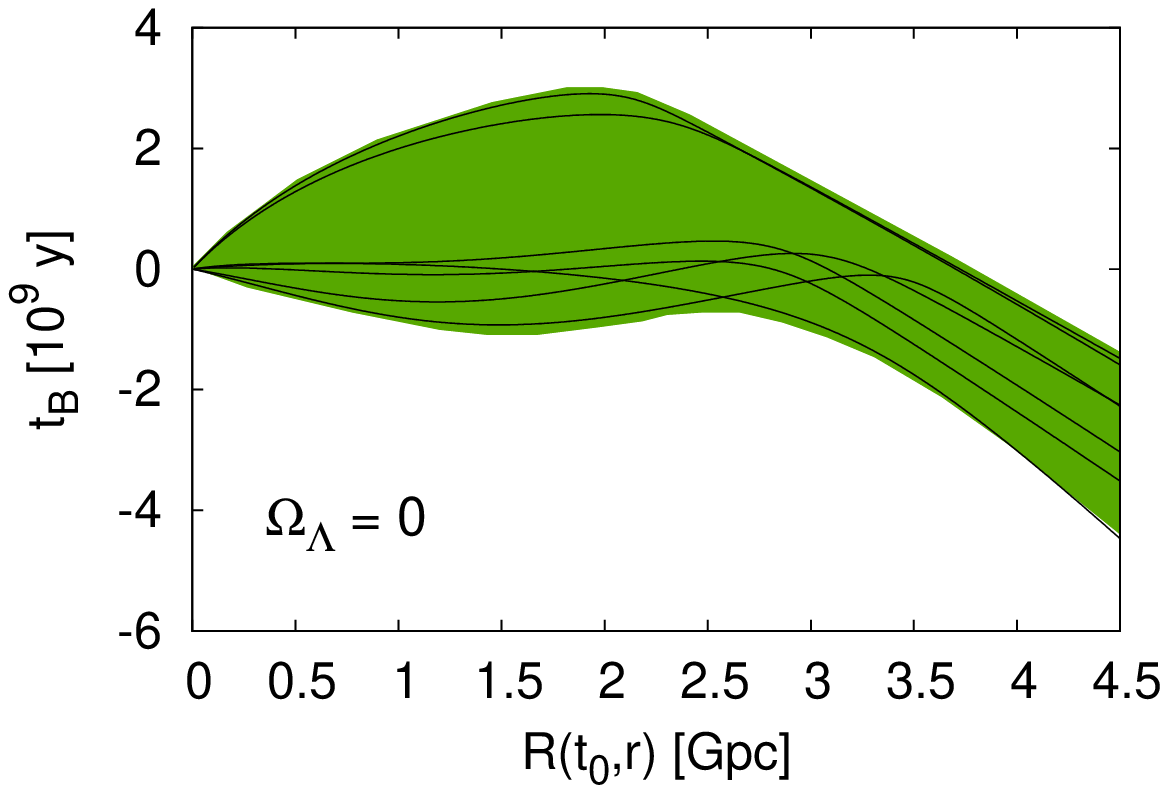}
 \includegraphics[scale=0.5]{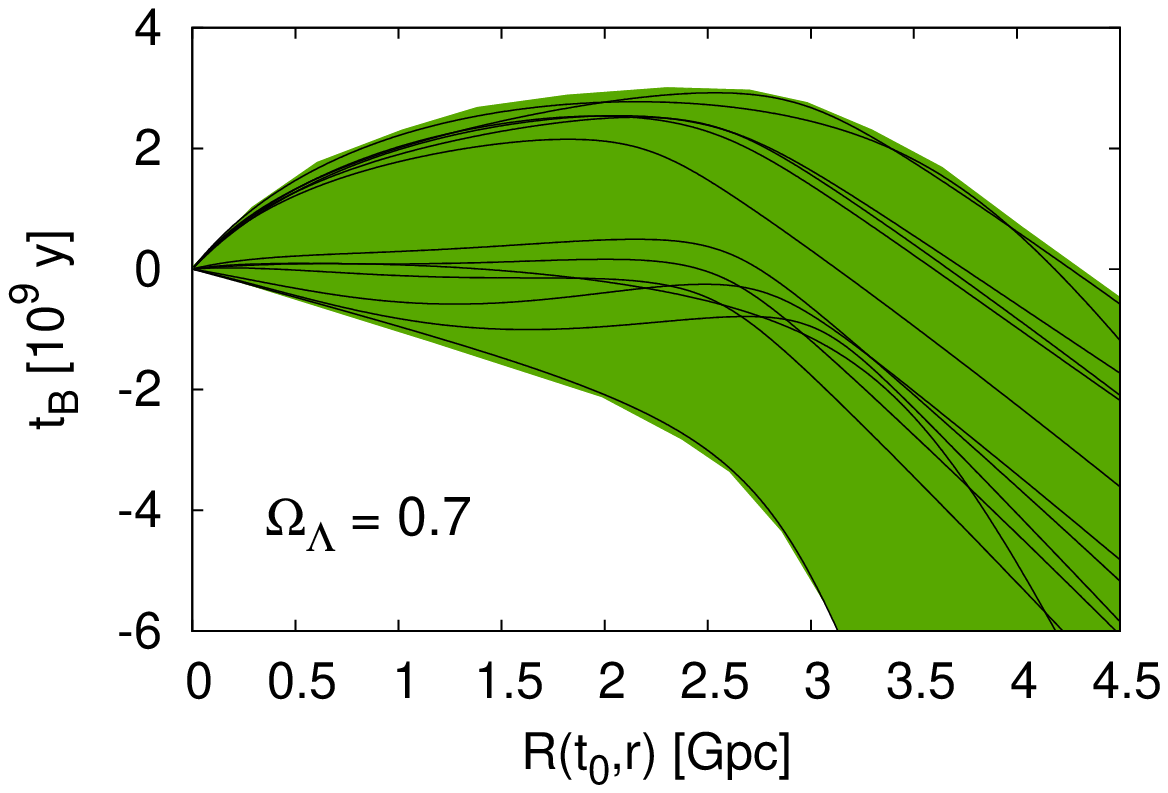}
 \caption{
 Constraints on the bang time function from luminosity distance and maximum stellar age data.
 The curves are the individual fits, and the green region is the envelope of all reasonable models.
 \label{fge2} }
 \end{center}
 \end{figure}

   The error band and the individual model curves for the present-day density profile are presented in Fig.\ \ref{fge1}, and those for the bang time function $t_B$ in Fig.\ \ref{fge2}.  These show that, for models with $\Lambda=0$, the dispersion in the $\Omega_m$ plot is not as large as in the case of $\Omega_\Lambda=0.7$.  Also, with $\Omega_\Lambda = 0$, a clear increasing trend is visible up to $R_0 = 3.5$ Gpc at least, whereas in the case of $\Omega_\Lambda=0.7$ there is far more uncertainty, and no discernible trend. 
  As to the bang time function, both cases allow for decreasing, increasing or a constant $t_B$ function up to at least $3.5$ Gpc.  

   The distance of 3.5 Gpc corresponds to $z \approx 1.5$, and at this distance the fit to the data is no longer reliable; as shown in Fig.\ \ref{figdata}, for $z > 1.5$ there is only small number of data.  Therefore the behaviour of these models at larger distances cannot be considered at all accurate, though it is interesting to notice that in all cases (with and without the cosmological constant) for $R_0>3.5$ Gpc there is a definite decreasing trend.

 \section{Conclusions}
 \label{conc}

   We have pursued the broad question of what observations can tell us about the geometry of our universe, and to what extent the observations support homogeneity.  As the cosmological datasets become larger, more accurate, and reach deeper $z$, we should be able to relax our assumptions and derive our spacetime metric in increasing detail, verifying to what extent, and on what scale, it is homogeneous on average.  While some assumptions will always be required, it is important to test different sets of assumptions, and to test theories in different ways.

   Here we have shown how measurements of maximum galactic ages, together with SNIa luminosities, over the available range of redshifts, can be used to determine the free metric functions of a spherically symmetric cosmology --- the \LT\ metric.  Previous work used galaxy number counts instead of maximum ages.  Reasons why a spherically symmetric model is a good place to start such an investigation, and appropriate to currently available datasets, are discussed above and elsewhere \cite{BKHC2009,LH07,MH08,Hel09}.  The algorithm does a numerical integration of the DEs we derived, but requires careful treatment of the origin, $z = 0$, and the neighbourhood of the maximum in the diameter distance, $\Rh = \Rh_m$ --- the apparent horizon, to avoid numerical difficulties.  We have adapted the apparent horizon correction method used for a different dataset combination \cite{Hel06,LH07,MH08} to our case, and shown how the properties of this apparent horizon provide relationships between observables and integrated quantities that allow one to detect a cumulative systematic error in the data and make a correction, thus ensuring a self-consistent solution.  All attempts to integrate past null cone data face potential divergences at the apparent horizon.  We have discussed the problem of extracting smooth functions from discrete data, and estimating the uncertainties in the output.  We adopted a method, based on a range of possible fitting functions, that is appropriate to the limited and noisy data currently available.  

   Also, in contrast to the inverse methods usually considered, we did not limit ourselves to the zero-$\Lambda$ LT models.  Usually, when considering inhomogeneous models, people are seeking to explain the data without the cosmological constant.  Recently however, inhomogeneous models with the cosmological constant have started to be considered more seriously; see for example \cite{MaPa2010} or \cite{BoCe2010}.  In order to fully test the assumption of homogeneity we cannot set $\Lambda=0$, thus we consider two cases $\Omega_\Lambda=0$ and $\Omega_\Lambda=0.7$.

   We have executed our algorithm using currently available data, and presented the results.  Though the data is not yet good enough to attach any confidence to our output, we have analysed our graphs as an interesting exercise.  In fact our results are quite reasonable, and in particular we have shown that there is no conflict between the data used and an LT model without $\Lambda$.  We expect much stronger constraints to emerge as the data improves.  It should be emphasised that the stated confidence levels in the concordance model are much higher exactly because the homogeneity assumption removes huge amounts of potential uncertainty.  Homogeneity is not being tested in those confidence calculations.  We note that comparisons of north and south values in e.g. the Sloan data, or angular variations in the Hubble flow, have yielded differences of uncertain significance.

   Thus we have demonstrated that the metric extraction method is both theoretically and numerically viable, and should be developed and extended, as an important and strong check on homogeneity.  It is because both time and distance vary down our past null cone in the radial direction, so observations are affected by source evolution as well as cosmic expansion, that testing radial homogeneity on top of that --- without assuming it somewhere --- is very difficult.  The relationship between the the paths of light rays --- the past null cone --- and the geometry (metric) they pass through is intrinsically non-linear.  Therefore our analysis of observations must take this into account.  The generalisation of the metric extraction method to handling angular variation is intrinsically complex, but must be pursued.
 
   In \cite{LLLW10} the authors took an LT model that reproduces the luminosity and density curves of the currently preferred \LCDM\ model.  They estimated the age of the universe at the central location (at us) for this LT model, and with their method they found the age of their chosen LT model near us is $< 11.7$~Gy at a $1 \sigma$ confidence level, but still within $1 \sigma$ of the locally measured age limits.  We point out that they did not find the best fit of an LT model to the observations.  They assumed the luminosity and density curves of the best fit FLRW model are the best fit for any model.  However, once one allows inhomogeneity, as one should if LT is being used, there is much more leeway in the luminosity and density data curves than is apparent from the best-fit FLRW parameters and the quoted confidence levels.  In contrast, our work uses the data to directly construct the model, and finds the age measurements are consistent with a range of reasonable inhomogeneous models (including some close to homogeneous).

 \ack 
 KB is grateful to the ACGC-UCT, where this research was initiated, for their hospitality.  
 KB also acknowledges the support of the Marie Curie Fellowship under the grant PIEF-GA-2009-252950.  
 CH thanks South Africa's National Research Foundation (NRF) for support.  
 AHAA thanks the Deutscher Akademischer Austausch Dienst (DAAD) for a bursary.

 \end{document}